# A Relativistic Gravitational Model Based on an Atom's Behavior in a Gravitational Field

Yehea I. Ismail

Northwestern University
Evanston, IL 60208
ismail@ece.nwu.edu

*Abstract -* **A simple general relativity theory for objects moving in gravitational fields is developed based on studying the behavior of an atom in a gravitational field and maintaining the principle of relativity. The theory complies with all the known effects of gravity such as the gravitational time dilation and faster light speeds higher in the gravitational field. The field equations are applied to calculate the satellite time dilation in any orbit, the light deflection by the sun, and the anomalous advance of Mercury's perihelion. In all these calculations, the results matched observations with an error of less than 1%. The approach to the new theory introduced here is different from the geometric approach used by the general relativity theory. The theory is field based where the potential energy of a system of masses can be easily calculated and the force can be found as the gradient of the potential field in analogy to Newtonian mechanics. The resulting field equations become the traditional Newton's equations when week gravitational effects are present. The special relativity theory of an object moving without experiencing gravitational fields can be derived directly from the gravitational field equations introduced here. The theory introduced here has several differences from the general relativity theory. For example, the event horizon of a black hole (where light cannot escape) has to be of zero radius, essentially meaning that light can escape any object unless the object has infinite density. Another primary consequence of this study is that the principle of equivalence of gravitational and inertial mass has limited validity and a new definition of gravitational mass is given here. Besides its extreme simplicity as compared to general relativity, the new theory removes all the known infinities and puzzles that can result from the general relativity theory and Newtonian mechanics. In addition, the new theory is in full compliance with quantum mechanical concepts and it is shown that the very essence of quantum mechanics is gravitational in nature and that electrons are gravitational black holes. Finally, a striking relation between the potential energy stored in the universe and the total mass energy of the universe flows naturally from the field equations introduced here which explains an observation that Feynman referred to as the great mystery.**

## I. Introduction

In this paper, a different approach from the geometric model used in general relativity [1]-[5] is followed to arrive at a relativistic gravitational field model. This new model is based on examining the behavior of an atom moving in a gravitational field and generalizing the observations deduced from this behavior. An atom is a basic structure of all materials and hence, the observations made by examining the behavior of an atom in a gravitational field can be extended to larger objects. In addition to the behavior of an atom in a gravitational field, the principle of relativity is maintained when deriving the field equations. Einstein's definition of the principle of relativity [1]-[5] is used here which states that all the physical laws and constants should be maintained in different frames of reference. Alternatively, if an experiment is performed and observed at a certain frame of reference, the results should be identical to when the experiment is repeated and observed in a different frame of reference. The definition of a frame of reference depends on the speed of the frame in special relativity [1], [4]. Another definition is given later in this paper for a frame of reference in a gravitational field as an equipotential surface based on an atom's behavior. Using this approach, simple field equations are derived that describe a complete relativistic gravitational model where all the effects of a distribution of masses in space [5]-[8] can be deduced, including motion, energy conservation, time dilation [9]-[13], light deflection [14]-[20], and orbit precession [21]-[23]. The new theory also removes the flows and contradictions that result from the general relativity theory, *e.g.*, [24]-[35].

A large amount of evidence is provided in the paper illustrating the validity of the new model and its compliance with most of the known physical effects of gravitation [5]-[8]. The first evidence is the way the model is derived. Starting by an atom and the principle of relativity and arriving at all the known effects of gravitation shows consistency in favor of the new approach. Calculations based on the model for satellite time dilation in any orbit [9]-[13], the light deflection by the sun[14]-[20], and the anomalous advance of Mercury's perihelion [21]-[23] matched observations with an error of less than 1%. Another evidence is that all the equations of the special theory of relativity can be directly derived from the new gravitational model as a special case despite the fact that the approach used to derive the gravitational model uses nothing from the special relativity theory. Other evidences come from the self-consistency of the new model. For example, the derived potential is such that if an observer at higher potential sees another observer at lower potential twice as slow, the lower observer sees the higher one twice as fast. Note that Newtonian-like potentials cannot support such a behavior as will be explained later. Another set of evidence comes from the physical and intuitive nature of the solution where many of the infinities and puzzles of gravitation are eliminated, *e.g.*, [24]-[35]. Also, the gravitational model is compatible with quantum mechanics and elementary point particles. It is actually shown that gravitation is the fundamental force controlling the interactions between photons and electrons.

In addition to the above evidences, one important evidence comes from the natural and successful explanation of a previously unexplained observation. By using the available estimates of the universe density and size, many people have pointed out a "spectacular coincidence" as Feynman refers to it in [26]. This observation is that the total potential energy stored in the universe is the same order of magnitude as the total energy stored in the mass of the universe (given by $Mc^2$, where $M$ is the total mass of the universe and $c$ is the speed of light) [26], [27]. The field equations introduced here show that these two quantities have to be *exactly equal* for any system of masses, not only for the universe because of some coincidence. The general relativity can provide no explanation of this fact and that is why Feynman referred to it as a great mystery (see page 10 of [26]).

The rest of the paper is organized as follows. In section II, the behavior of an atom in a gravitational field and the concept of relativity are used to deduce the physical laws governing gravitation. In section III, the field equations are derived and shown to reduce to Newton's equations at low gravitation. The energy conservation and deduction of the special relativity equations from the gravitational model are also discussed in section III. The field equations are applied to calculate the satellite time dilation in any orbit, the light deflection by the sun, and the anomalous advance of Mercury's perihelion in section IV and the results matches observations with a relative error of less than 1%. Some physical implications of the new gravitational model are discussed in section V including the "great mystery" and relations to quantum mechanics. In section VI, the fundamental differences in the assumptions underlying this theory and those underlying general relativity are stressed. Several important questions are considered in section VI such as why does general relativity seem to work (essentially in week gravitation), where it fails, and what physical contradictions can general relativity cause. Finally, conclusions are given in section VII.

## II Physical Concepts Underlying Gravitation

The physical concepts and laws governing gravitation are established in this section. The behavior of an atom in a gravitational field is discussed in subsection A from which the gravitational time dilation and frequency shifts can be directly deduced. In subsection B, it is shown that the principle of relativity implies space dilation in association with gravitation. This space dilation is shown to be necessary for compliance with the known results of the varying speed of light in gravitational fields which causes light to fall into gravitational fields. In subsection C, a major contradiction between the principle of equivalence of inertial and gravitational masses [2] and energy conservation is pointed out and hence the principle of equivalence has to be discarded. Note that the fact that gravitational and inertial mass are not generally equal does not contradict well known experiments [36]-[43] that illustrate their equivalence as will be explained in subsection C. A definition of gravitational mass is thus given and its relation to energy and inertial mass are discussed.

### A. Behavior of Atoms in Gravitational Fields

In this section, Bohr's model of a hydrogen atom [44], [45] is used to investigate the effects of gravitation. A hydrogen atom has one electron orbiting a heavy nucleus with one proton and one neutron as shown in Fig. 1. According to Bohr's model, the electron is allowed only specific orbits related to the energy of the electron. The radii of these orbits are given by

$$r_n = \frac{n^2 h^2 \varepsilon_o}{\pi m_e e^2}, \qquad (1)$$

where $n$ is an integer taking the values 1, 2, 3, ... and $r_n$ is the $n^{th}$ allowed electron orbit with $n = 1$ representing the orbit closest to the nucleus. The mass and charge of an electron are given by $m_e$ and $e$, respectively. $h$ is Plank's constant and $\varepsilon_o$ is the permittivity of vacuum. The velocity of the electron at the $n^{th}$ orbit is given by

$$v_n = \frac{nh}{2\varepsilon_o m_e r_n}. \qquad (2)$$

The kinetic energy of the electron in the $n^{th}$ orbit is given by

$$E_n = \frac{-e^2}{8\pi\varepsilon_o r_n} = \frac{-m_e e^4}{8n^2 h^2 \varepsilon_o^2}. \qquad (3)$$

For an electron to move from one orbit to an adjacent orbit, the electron has to absorb or emit a photon carrying the amount of energy given by the energy difference of the two orbits. Hence, for an electron to move from the second to the first orbit, the electron has to emit a photon with a frequency given by

$$f_{ph} = \frac{E_2 - E_1}{h}, \tag{4}$$

since the energy of a photon is given by the well known formula $E_{ph} = hf_{ph}$.

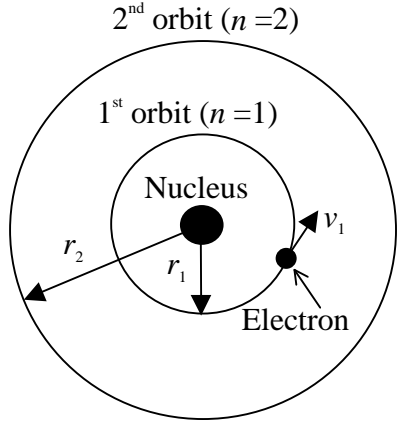

Fig. 1. Bohr's model of a hydrogen atom. Note that this qualitative figure is not drawn to scale.

Now consider moving an atom from a low position close to the earth surface to a higher position (say into a satellite). The atom gains gravitational potential energy which can be retained as kinetic energy if the atom freely falls under the effect of gravity. But where exactly does this energy go and how is it distributed among the different energies constituting the atom. A good point to start with is to consider what happens to the constituent particles (such as electrons and protons) of the atom as they are raised individually against gravitation. The *inertial* mass of a particle of mass $m$ at earth increases to $m+E_{pot}/c^2$ where the famous formula $E = mc^2$ has been employed and $E_{pot}$ is the potential energy gained by the mass $m$. Hence, the mass of the electron at the higher position is given by

$$m_{eup} = \lambda m_{edown}, \tag{5}$$

where

$$\lambda = 1 + \frac{E_{pot}}{m_{edown} c^2}, \tag{6}$$

Note that $\lambda$ is independent of the mass of the object and only depends on the starting and ending positions of the object. This result is due to the fact that $E_{pot}$ gained by any object is proportional to the mass of this object as will be illustrated later. Note also that $\lambda$ represents the increase in the total energy of the body as it moves from a lower to a higher potential since the total energy of the body is given by $mc^2$. By substituting (5) into Bohr's atom model, the following relations result

$$r_{up} = \frac{r_{down}}{\lambda}, \quad v_{up} = v_{down}, \quad E_{up} = \lambda E_{down}, \quad \text{and} \quad f_{phup} = \lambda f_{phdown}. \tag{7}$$

The radius of the atom shrinks by $\lambda$, the velocity of the electron in any orbit remains constant, the binding energy of the electrons increases by $\lambda$, and the photons emitted by the atom are blue shifted by a factor of $\lambda$. Note that the number of revolutions of the electron around the atom increases by a factor of $\lambda$ since the speed remains constant while the orbit radius shrinks. Thus, the potential energy gained by the whole atom is distributed uniformly among the masses of the constituting particles, the binding energy, and the emitted or absorbed photons, with each type of energy increasing by a factor of $\lambda$. Bohr's atom model guarantees both the conservation of energy and the uniformity of energy distribution in the atom, which makes physical sense since there is no reason the added potential energy should concentrate in one type of energy rather than others. Note also that an emitted photon travelling with the speed of light would travel the distance occupied by a number of atoms that increases by $\lambda$ at high positions as compared to earth since the size of each atom has shrunk by $\lambda$.

Several observations in agreement with the known effects of gravitation can be made based on the above discussion. For someone observing the high atom from earth, he will notice that the number of revolutions of the electron around the atom has increased in relation to a similar atom on earth. If he takes that as a measure of time, he can interpret this observation as time is running faster at the higher potential. Also, he will notice that the photons emitted by the higher atom are blue shifted relative to the photons emitted by an atom on earth. This blue shift is a well-known effect of gravitation [9]-[13], [26]. To see what an observer at the same height as the atom notices, assume $\lambda = 2$. This observer will measure that a photon has traveled between two points a distance given by twice that he would measure at earth since all his measuring rods (made of atoms) shrunk in size. However, he also measures the time taken to cut the distance between the two points as doubled since his clock works twice as fast. Hence, the higher observer will measure the same speed of light and won't feel any change in the physical laws as required by the principle of relativity. He would also shrink in size so he cannot tell that anything has shrunk around him.

**B. Space Dilation**

The above study of the behavior of an atom in a gravitational field shows very good agreement with the known effects of gravitation and the principle of relativity which states that someone moving from one frame of reference to another frame of reference should experience the same physical laws. Alternatively, a person should have absolutely no way to tell that he has changed his frame of reference by making any experiment in his frame of reference. Now is a good point to make a more precise definition of a frame of reference for stationary (or low speed) objects in a gravitational field. A frame of reference in a gravitational field is defined here as an equipotential surface. This definition is due to the fact that it is potential energy that changes the time and atom sizes in such a way that any observer in the same potential will not feel any change in the physical laws. However, an observer at a different potential would see changes in the physical laws and quantities when observing an object that changes its position moving between different potentials.

However, there is a serious violation of the principle of relativity. Consider a person measuring the distance between two objects in his frame of reference using his rulers. The two objects have vacuum in between (see Fig. 2). As he moves to a higher potential, his rulers shrink. If we assume that space does not shrink he will measure more distance as he moves to a higher potential and thus be able to tell that he has changed his frame of reference in direct violation to the principle of relativity. In fact there are numerous experiments that will violate the principle of relativity if space does not shrink.

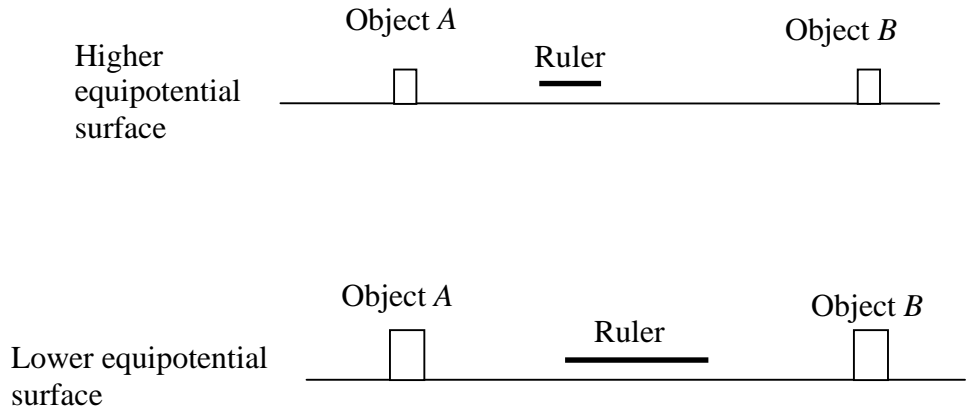

Fig. 2. Distance measurement between objects at different frames of reference if space does not shrink.

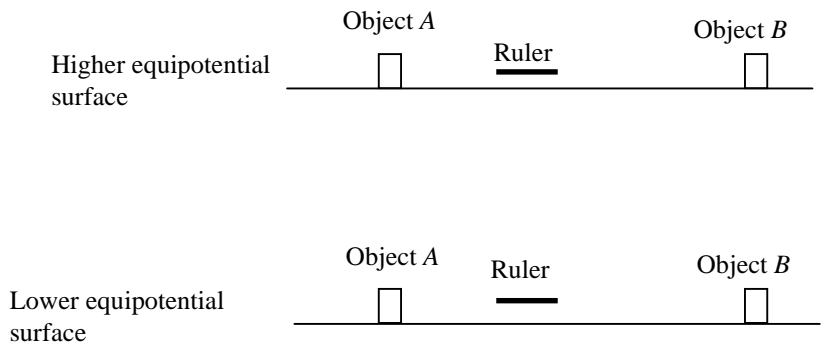

Fig. 3. Distance measurement between objects at different frames of reference if space shrinks. The image given here is as observed by someone at the higher frame of reference.

Hence, the space should shrink with a factor $1/\lambda$, which is the same as the shrinking factor of an atom to maintain the principle of relativity. In this case, an observer moving from a lower potential to a higher one will measure exactly the same distance between any two points in his frame of reference. Note also that this space shrinking is *omni-directional* since the atom shrinks in size in all directions. Moreover, the higher observer will not perceive an object in his frame smaller than a similar object at a lower potential since the whole space has shrunk from his point of view including the lower objects. See Fig. 3. Hence, as objects change their frame of reference they always perceive similar objects at other frames as having the same size. Alternatively, the size of objects at any frame of reference appears the same no matter which frame of reference they are observed from. The same comments apply for distances between objects.

This space shrinking is also necessary for the conservation of energy in electromagnetic fields. For example consider two charges that move form a lower potential to a higher one. The distance between the two charges shrinks by a factor of $1/\lambda$. Hence, the force between the two charges increases as $\lambda^2$ since the electromagnetic force has an inverse square dependence on the distance between charges. The potential energy between the two charges is given by the integration of $Fdr$, where $F$ is the force between the two charges and $dr$ is an incremental distance between two points. Hence, the potential energy increases linearly with $\lambda$ as required by the law of conservation of energy. Also, the quadratic increase with $\lambda$ of the force between charges will make all the movements of the charges faster in proportion to $\lambda$ which is required by the principle of relativity since all the clocks also run faster.

Now we have to update our view of how the lower observer will observe the changes in an atom's behavior as it is raised to a higher gravitational potential. The lower observer will see the atom having the same size as on earth. He will still see the electron making a larger number of revolutions around the nucleus per second, which he will interpret as the electron is going faster, and he will see the emitted photons blue shifted. Most importantly, he will see the speed of light at the higher potential as larger (quadratically with $\lambda$) than the speed of light on earth. This is a well-known fact and causes the light to fall (or bend) under the effect of gravity [14]-[20]. This is the observation that initiated Einstein's interest in gravitation and led to the theory of general relativity. Hence, in summary, the lower observer will see all the objects at a higher potential having the same size and will see everything at the higher potential faster than in his frame, including the speed of light. A higher observer will see all the processes slower at the lower potential. Also, all the physical laws will be maintained no matter what frame of reference an experiment is conducted given that the results are observed from the same frame of reference.

**C. Inertial and Gravitational Mass**

So far, it was shown that all the phenomena of gravitation can be deduced from the behavior of an atom in a gravitational field and by maintaining the principle of relativity. If someone did not know any of the gravitational effects he could have deduced them in the manner described in the preceding sections. However, the effect of gravitation on one more process is not discussed yet. That is the effect of gravitational potential on gravitation. To illustrate what this means, consider two small objects of masses $m_1$ and $m_2$ moving from one frame of reference to another in the field of a much larger mass $M$. See Fig. 4. An observer in the same frame of reference of the two masses will measure a force between the two masses given by Newton's law as

$$F = G\frac{m_1 m_2}{r^2}, \qquad (8)$$

where $r$ is the distance between the two bodies and $G$ is the gravitational constant as defined in Newton's law. The potential energy change in this system of two masses as the distance between the two of them change from $r_1$ to $r_2$ is given by

$$\Delta E_{pot} = Gm_1 m_2 \left[\frac{1}{r_2} - \frac{1}{r_1}\right]. \qquad (9)$$

The acceleration on of $m_2$ at any radius $r < d$ when $m_1$ is held in position is given by

$$a = G\frac{m_1}{r^2} = \frac{k}{r^2}, \qquad (10)$$

where $k = Gm_1$. Finally, if the two masses start at rest at a distance $d$ from each other and are only acted upon by the gravitational force between them in the direction of the line connecting the radii, the speed of $m_2$ at any radius $r < d$ when $m_1$ is held in position is given by

$$v = \sqrt{2k\left[\frac{1}{r} - \frac{1}{d}\right]}. \qquad (11)$$

Now if the two masses are moved to a higher potential, and if the gravitational mass is assumed equal to the inertial mass, then both $m_1$ and $m_2$ should increase in all the above equations by a factor of $\lambda$. Also, as discussed in the previous section, space dilation is necessary to maintain the principle of relativity. Hence, the space between the two masses shrinks by a factor of $1/\lambda$. These transformations result in the following relations

$$F_{up} = \lambda^4 F_{down}, \quad E_{up} = \lambda^3 E_{down}, \quad a_{up} = \lambda^3 a_{down}, \quad \text{and} \quad v_{up} = \lambda^2 v_{down}. \qquad (12)$$

These relations directly violate the law of conservation of energy since all the energies should increase by a factor of $\lambda$. Also, the principle of relativity is violated since the time at which the two particles collide will decrease quadratically with $\lambda$. Hence, an observer making experiments involving gravitation and other forces such as the electromagnetic force or atomic processes would find that the experiment changes as he moves to a higher position. If he is using an atomic clock to measure the time at which the two particles collide, he will find this time decreasing as he moves to a higher potential since his clock speeds up linearly with $\lambda$ while the collision time speeds up quadratically with $\lambda$. Also, the ratio between the electromagnetic and gravitational forces changes as the frame of reference changes. Hence, all the physics and experiments involving gravitation are scrambled if performed at higher.

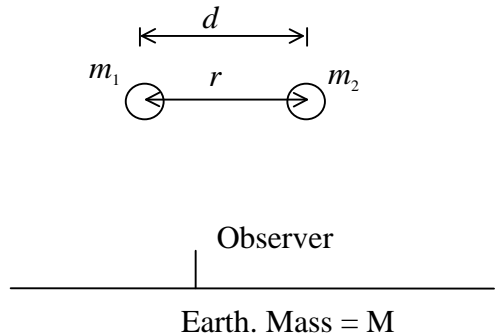

Fig. 4. Two small objects of masses $m_1$ and $m_2$ moving from one frame of reference to another in the field of a much larger mass $M$.

A way to resolve all these inconsistencies is to question the principle of equivalence. Assume two different types of masses, the gravitational mass which determines the gravitational force and the inertial mass which determines the acceleration due to any type of force. Also, it is assumed that the energy does not add to the gravitational mass while it adds to inertial mass. The gravitational mass is differentiated from the inertial mass by adding a *g* suffix to the mass. Hence, $m_1$ is the inertial mass while $m_{1g}$ is the corresponding gravitational mass. Under these assumptions, equations (8)-(11) become

$$F = G \frac{m_{1g} m_{2g}}{r^2}, \tag{13}$$

$$\Delta E_{pot} = G m_{1g} m_{2g} \left[ \frac{1}{r_2} - \frac{1}{r_1} \right]. \tag{14}$$

$$a = G \frac{m_{g1}}{r^2} = \frac{k}{r^2}, \tag{15}$$

$$v = \sqrt{2k \left[ \frac{1}{r} - \frac{1}{d} \right]}. \tag{16}$$

Using these relations, as the experiment is performed at a higher potential, the following transformations in the gravitational force, potential energy, and speed result

$$F_{up} = \lambda^2 F_{down}, \quad E_{up} = \lambda E_{down}, \quad a_{up} = \lambda^2 a_{down}, \quad \text{and} \quad v_{up} = \lambda v_{down}. \tag{17}$$

Note that the gravitational mass remains constant at higher and lower positions since energy is assumed not to add to gravitational mass. These results are in full compliance with energy conservation and with the principle of relativity. The gravitational force, electromagnetic force, and all other forces scale quadratically with $\lambda$. Also, the time it takes for the two masses to collide decreases by $\lambda$ since the speed increases by $\lambda$. Hence, all the physical laws and energy conservation are maintained under this model.

But what about all the experiments that have been performed to detect any difference between the gravitational and inertial mass [36]-[43] and concluded to a very high degree of accuracy that these masses are equal. The answer is that all these experiments can never detect any difference since the experiments are performed and the results are observed in the same frame of reference. A person will never be able to detect any difference between the gravitational and inertial mass in his frame of reference. To clarify this statement, consider some one measuring the two masses once on earth and another at a higher position. As the experiment is moved to a higher position, the gravitational mass remains constant while the inertial mass increases by a factor of $\lambda$. However, when performing experiments to measure the inertial mass in the higher position, the same value will result as that measured on earth because the inertial mass has increased but so did all the forces and energies. The changes are always such that the same value for inertial mass is always measured, which is actually required by the principle of relativity. This behavior is in analogy to the speed of light. The speed of light will always be measured the same no matter what frame of reference the experiment is performed in. However, when observing the speed of light from a different frame of reference, the value changes. Einstein's thought experiments about gravitational and inertial mass as well as all the experiments [36]-[43] performed to detect any difference between these two masses were performed in the same frame of reference. Note also that objects will fall at the same speed under gravity even with different inertial and gravitational masses since in a given frame of reference the two masses are equal. This fact is clear from the field equations derived later in the paper based on the above discussion. Hence, the principle of equivalence, on which the general relativity theory is based, has limited validity. Also, as will be discussed in section VI, the principle of equivalence leads to violation of energy conservation in atoms based on the quantum mechanical model of an atom and causes non-uniform scaling of the physical forces.

The following definitions and rules can be made concerning gravitational and inertial mass:
1. The gravitational mass of a body is the mass measured by any kind of experiment in any frame of reference.
2. The gravitational mass determines the force of gravitation.
3. This mass remains the same when viewed from any frame of reference.
4. The previous point can also be restated as: energy does not add to gravitational mass. Gravitational mass is conserved in the same way the charge is conserved [1], [46]-[49] when a body moves between frames of reference.
5. The inertial mass is the mass determining the acceleration of a body when acted on by forces.
6. Energy (from potential or motion) adds to this mass.
7. Thus, this mass appears different when viewed from different frames of reference.
8. The inertial and gravitational masses are always equal when measured in the same frame of reference.

### III. Relativistic Gravitational Field Model

In this section, simple relativistic field equations are derived based on the physical characteristics of gravitation that were discussed in the previous section. The potential field equations are first derived for the special case of two masses in subsection A where the basic characteristics of the new potential field are discussed. The potential field equations are extended to multi-mass systems in subsection B. In subsection C, the type of kinetic energy that is conserved under the new potential field is derived and the special relativity equations are shown to naturally result from the potential field. The equations governing the motion of two masses under the effect of gravity are discussed in subsection D. Finally, the general equations governing the motion of any number of masses under the effect of gravity are discussed in subsection E.

## A. Potential Field Equations for a System of two Masses

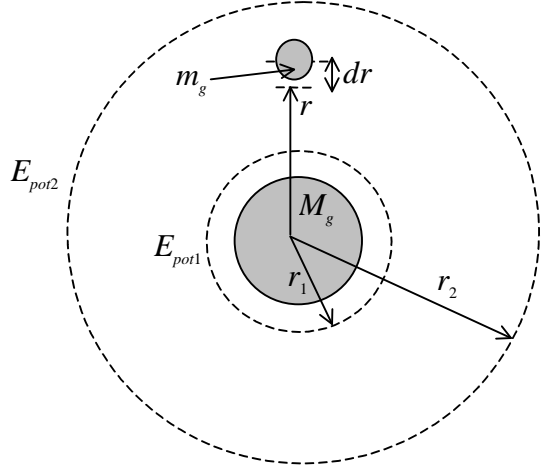

Fig. 5. An object of mass $m_g$ is moved from a distance $r_1$ to a larger distance $r_2$ in the gravitational field of another mass $M_g$.

   Consider a mass $m_g$ that is moved from a distance $r_1$ to a larger distance $r_2$ in the gravitational field of another mass $M_g$ as shown in Fig. 5. The body gains potential energy due to moving it against the gravitational force. This potential energy can be found by integrating the incremental energy gained by $m_g$ when the body is moved a small distance $dr$ at a distance $r$ where $r$ varies from $r_1$ to $r_2$. However, this potential energy has to be referred to some frame of reference since each observer measures energy differently according to his frame of reference. An observer at $r_1$ would measure the same amount of energy higher by $\lambda$ as compared to what an observer at $r_2$ would measure. When referred to $r_1$, the force acting on $m_g$ at a radius $r$ is given by

$$F(r) = G \frac{M_g m_g}{r^2} \lambda(r)^2. \tag{18}$$

The reason the $\lambda^2$ factor appears in the above equation is that $m_g$ sees a space that is shrunk at a radius $r$ by a factor of $\lambda$ with respect what an observer at $r_1$ sees and hence, $m_g$ feels itself closer to $M_g$ by a factor of $\lambda$, increasing the force quadratically. The incremental energy added to the body when moving a distance $dr$ is given by $F(r)dr/\lambda$ which is given by

$$dE_{pot}(r) = G \frac{M_g m_g}{r^2} \lambda(r) dr. \tag{19}$$

The division by $\lambda$ is because the distance $dr$ as seen by an observer at $r_1$ appears smaller by a factor of $\lambda$ in the reference frame of $m_g$ due to the space shrinking. Since $E_{pot1}$ is taken as the reference, $E_{pot1}$ can be set to zero which results in $\lambda(r)$ given by

$$\lambda(r) = 1 + \frac{E_{pot}(r)}{m_g c^2}, \qquad (20)$$

according to the discussion in section II. Hence, the total potential energy gained by $m_g$ when moving form $r_1$ to $r_2$ in the reference frame of $r_1$ can be calculated by integrating (19) as

$$\int_0^{E_{pot21}} \frac{dE_{pot}(r)}{\left(1 + \frac{E_{pot}(r)}{m_g c^2}\right)} = \int_{r1}^{r2} G \frac{M_g m_g}{r^2} dr. \qquad (21)$$

Hence, the potential energy in reference frame of $r_1$ is given by

$$E_{pot21} = m_g c^2 \left[\exp\left(\frac{GM_g}{c^2}\left(\frac{1}{r_1} - \frac{1}{r_2}\right)\right) - 1\right]. \qquad (22)$$

The factor $\lambda_{21}$ describing the scaling of physical quantities (time dilation, space dilation, energies, …) at $r_2$ as observed form $r_1$ can be found by substituting (22) in (20) and is given by

$$\lambda_{21} = \exp\left(\frac{GM_g}{c^2}\left(\frac{1}{r_1} - \frac{1}{r_2}\right)\right) \qquad (23)$$

Note that $\lambda_{21}$ take values between 1 and ∞. Hence, the time runs faster at $r_2$ by a factor of $\lambda_{21}$ and the space shrinks by a factor of $1/\lambda_{21}$ in reference to $r_1$. Also all the energies at $r_2$ increase by the same factor $\lambda_{21}$ as discussed in the previous section.

To examine the alternate view of what an observer at $r_2$ measures as the potential energy lost when $m_g$ moves from $r_2$ to $r_1$, $E_{pot2}$ is set to zero and the integration in (21) is repeated as

$$\int_0^{E_{pot12}} \frac{dE_{pot}(r)}{\left(1 + \frac{E_{pot}(r)}{m_g c^2}\right)} = \int_{r2}^{r1} G \frac{M_g m_g}{r^2} dr, \qquad (24)$$

which results in a potential energy loss as measured from the reference frame of $r_2$ given by

$$E_{pot12} = -m_g c^2 \left[1 - \exp\left(-\frac{GM_g}{c^2}\left(\frac{1}{r_1} - \frac{1}{r_2}\right)\right)\right]. \qquad (25)$$

The factor $\lambda_{12}$ describing the scaling of physical quantities at $r_1$ as observed form $r_2$ is given by

$$\lambda_{12} = \exp\left(-\frac{GM_g}{c^2}\left(\frac{1}{r_1} - \frac{1}{r_2}\right)\right) = \frac{1}{\lambda_{21}}. \qquad (26)$$

Note that $\lambda_{12}$ take values between 1 and 0. Hence, the time slows at $r_1$ by a factor of $\lambda_{12}$ and the space expands by a factor of $1/\lambda_{12}$ in reference to $r_2$. Also, all the energies at $r_1$ decrease by the same factor $\lambda_{12}$ in reference to $r_2$.

The above potential energy equations have very interesting characteristics. First, note that both equations tend to the potential energies given by Newton's law at low gravitation or mathematically when

$$\left( \frac{GM_g}{c^2} \left( \frac{1}{r_1} - \frac{1}{r_2} \right) \right) \ll 1. \tag{27}$$

If this condition is satisfied, (22) and (25) tend to

$$E_{pot21} = GM_g m_g \left( \frac{1}{r_1} - \frac{1}{r_2} \right) \quad \text{and} \quad E_{pot12} = -GM_g m_g \left( \frac{1}{r_1} - \frac{1}{r_2} \right), \tag{28}$$

which are simply the Newtonian potential energy change when $m_g$ moves form $r_1$ to $r_2$ and from $r_2$ to $r_1$, respectively. This is a very important feature since Newton's law has proved exceptional success at low gravitational effects.

Another important feature is that the dilation factors $\lambda_{21}$ and $\lambda_{12}$ are the reciprocal of each other. This is a crucial feature for the solution to be physically meaningful. This characteristic indicates that if an observer at $r_1$ sees processes running twice as fast at $r_2$, then an observer at $r_2$ sees processes running twice as slow at $r_1$, which is a physical requirement. The same can be said about space dilation (which maintains the same sizes of objects when perceived from different frames) and all types of energy. Note also that the maximum time and space dilations are infinity and the minimum is zero. The infinite value results only if there is some mass inside a radius $r_1 = 0$, which necessarily implies an infinite density. No negative or imaginary quantities are allowed. This is also a physically intuitive result since the theory can produce no infinities with masses of finite density. Also, no negative or imaginary time or journeys to the past are allowed. Note also that the reciprocal feature between $\lambda_{21}$ and $\lambda_{12}$ is not a likely coincidence since many other potential fields will simply not produce this result. For example, a Newtonian potential would have $E_{pot12} = -E_{pot21}$ in which case $\lambda_{21}$ and $\lambda_{12}$ are given by

$$\lambda_{21} = 1 + \frac{E_{pot21}}{m_g c^2} \quad \text{and} \quad \lambda_{12} = 1 - \frac{E_{pot21}}{m_g c^2}. \tag{29}$$

Obviously, $\lambda_{21}$ is not equal to $1/\lambda_{12}$. In the new model, the potential energy measured by an observer at $r_2$ is not negative of that measured by an observer at $r_1$. The two energies have different magnitudes as given by (22) and (25). These different magnitudes are such that the reciprocal relation between $\lambda_{21}$ and $\lambda_{12}$ is guaranteed. Note also that these different magnitudes does not imply any physical inconsistency since each observer measures the energy in reference to the energy level of his frame of reference which differ between $r_2$ and $r_1$ by a factor of $\lambda_{12}$. Equation (25) can be deduced from (22) by multiplication with $-\lambda_{12}$. The negative sign is due to the fact that energy is lost in one direction and gained in the other.

Note also that if the potential at $r_2$ in reference to $r_1$ is given by $E_{pot21}$, and the potential at some other radius $r_3$ in reference to $r_1$ is given by $E_{pot31}$, the potential of $r_3$ in reference to $r_2$ is always such that $\lambda_{32}\lambda_{21}=\lambda_{31}$. This characteristic can be verified by using (22) once with $r_1$ and $r_2$, another time with $r_1$ and $r_3$, and a third time with $r_3$ and $r_2$. This feature also has a physical meaning since if an observer at $r_1$ sees the physical processes running two times faster at $r_2$ and four times faster at $r_3$, then an observer at $r_2$ will see the physical processes at $r_3$ running twice as fast. Note that $E_{pot31}$ is not equal to $E_{pot32}-E_{pot21}$ as in the case of Newtonian mechanics. The correct potential should be found from the relation $\lambda_{32}=\lambda_{31}/\lambda_{21}$.

Another important feature is that the equipotential surfaces around a spherical mass are still spheres and the value of the potential depends only on the radius of the sphere in analogy to Newtonian potentials. However, the dependence on the radius is exponential in this solution while the dependence is proportional to $1/r$ in the case of Newtonian potentials. Note that the potential energy as measured by an observer at $r_2$ is bound to a minimum of $-m_g c^2$ at $r_1 = 0$. Note that the Newtonian potential becomes negative infinity at $r_1 = 0$, which leads to some nonphysical conclusions as discussed later. This difference has large physical implications that will be discussed in more detail in section V.

Finally, note that the field equations in (22) and (25) are conservative and the integration of $\vec{F} \cdot \vec{dl}$ over any closed path is equal to zero. This characteristic can be easily proven using the traditional graphical interpretation used in Newtonian fields and in electromagnetic fields and noticing that to make a closed loop the object has to cross each frame of reference twice, once with the gravitational force in the same direction as *dl* and another with the gravitational force in opposite direction to *dl*. Hence, the whole integration reduces to zero. This is an important feature indicating that energy is conserved under this gravitational model. The energy conservation will be discussed in more detail in subsection C.

## B. Potential Field Equations for Multiple Mass Systems

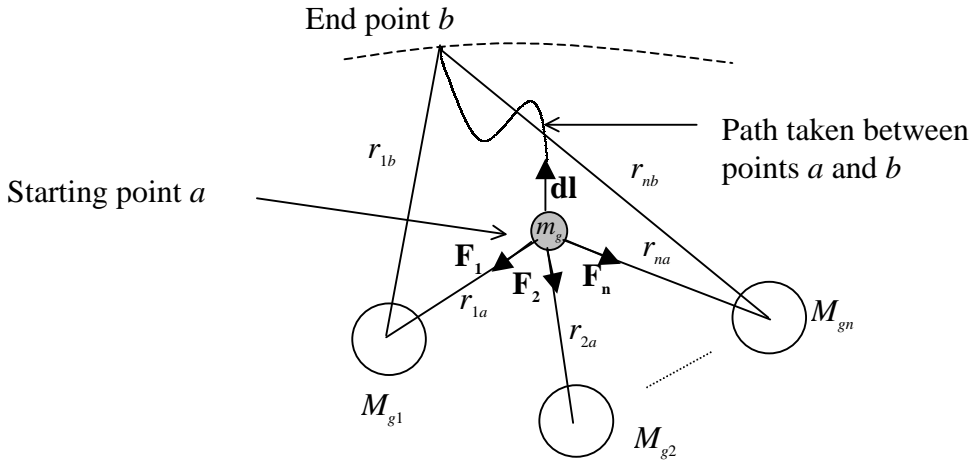

Fig. 6. An object of mass $m_g$ is moved from point *a* to point *b* in the gravitational field of a set of static masses $M_{g1}$ - $M_{gn}$.

Consider the distribution of masses shown in Fig. 6. An object of mass $m_g$ is moved from point *a* to point *b* in the gravitational field of a set of static masses $M_{g1}$ - $M_{gn}$. The incremental potential energy gained by $m_g$ as it moves a distance **dl** in the reference frame of an observer at point *a* is given by

$$dE_{pot} = m_g \left[ \sum_{i=1}^{n} G \frac{M_{gi}}{r_i^2} \hat{\mathbf{r}}_\mathbf{i} \right] \cdot \mathbf{dl} \left( 1 + \frac{E_{pot}}{m_g c^2} \right), \quad (30)$$

where a bold vector notation is used and $r_i$ is the distance between $m_g$ and $M_{gi}$. The dot product in (30) is simply given by the component of **dl** in the direction of $r_i$ denoted $dr_i$. Hence, (30) can be integrated as

$$\int_0^{E_{pot\_ba}} \frac{dE_{pot}}{\left(1 + \frac{E_{pot}}{m_g c^2}\right)} = m_g \left[ \sum_{i=1}^{n} \left( \int_{r_{ia}}^{r_{ib}} G \frac{M_{gi}}{r_i^2} dr_i \right) \right], \quad (31)$$

where the potential energy at $a$ was taken as zero. Hence, the potential energy when moving $m_g$ from $a$ to $b$ in the reference frame of an observer at $a$ is given by

$$E_{pot\_ba} = m_g c^2 \left[ \exp\left( \sum_{i=1}^{n} \frac{GM_{gi}}{c^2} \left( \frac{1}{r_{ia}} - \frac{1}{r_{ib}} \right) \right) - 1 \right]. \quad (32)$$

Similarly, the potential energy of $a$ relative to an observer at point $b$ is given by

$$E_{pot\_ab} = -m_g c^2 \left[ 1 - \exp\left( -\sum_{i=1}^{n} \frac{GM_{gi}}{c^2} \left( \frac{1}{r_{ia}} - \frac{1}{r_{ib}} \right) \right) \right]. \quad (33)$$

The time dilation factors from the points of view of $a$ and $b$ are given by

$$\lambda_{ab} = \exp\left( -\sum_{i=1}^{n} \frac{GM_{gi}}{c^2} \left( \frac{1}{r_{ia}} - \frac{1}{r_{ib}} \right) \right) = \frac{1}{\lambda_{ba}}. \quad (34)$$

The solution exhibits all the characteristics of the two mass solution discussed in subsection *A*. The potential energy tends to Newton's expressions at low gravitation when the argument of the exponential is much less than one. The reciprocal property between $\lambda_{ab}$ and $\lambda_{ba}$ is maintained. The field is still conservative and $\lambda_{ab}$ is bound between one and zero while $\lambda_{ba}$ is bound between one and infinity guaranteeing a physical time and space dilation.

The potential energy acquired by $m_g$ when moving from $a$ to $b$ as viewed by a general observer at point $o$ can be easily verified to be given by

$$E_{pot\_ba} = m_g c^2 \left[ \exp\left( \sum_{i=1}^{n} \frac{GM_{gi}}{c^2} \left( \frac{1}{r_{ia}} - \frac{1}{r_{ib}} \right) \right) - 1 \right] \exp\left( \sum_{i=1}^{n} \frac{GM_{gi}}{c^2} \left( \frac{1}{r_{io}} - \frac{1}{r_{ia}} \right) \right) \quad (35)$$

where $r_{io}$ is the distance between the point $o$ and the mass $M_{gi}$. The extra exponential in (35) as compared to (32) is simply $\lambda_{oa}$ since an observer at $o$ measures all energies scaled by $\lambda_{oa}$ as compared to an observer at $a$. The last expression can be put in vector notation as (see Fig. 7)

$$E_{pot\_ba} = m_g c^2 \left[ \exp\left( \sum_{i=1}^{n} \frac{GM_{gi}}{c^2} \left( \frac{1}{|\mathbf{r_i} - \mathbf{r_a}|} - \frac{1}{|\mathbf{r_i} - \mathbf{r_b}|} \right) \right) - 1 \right] \exp\left( \sum_{i=1}^{n} \frac{GM_{gi}}{c^2} \left( \frac{1}{|\mathbf{r_i}|} - \frac{1}{|\mathbf{r_i} - \mathbf{r_a}|} \right) \right) \quad (36)$$

This vector formulation gives the minimum number of variables given by the distances form the point $o$ to all the masses $M_{gi}$ (denoted $\mathbf{r_i}$) and to the points $a$ and $b$ (denoted $\mathbf{r_a}$ and $\mathbf{r_b}$), a total of $n + 2$ vector variables. This formulation will become necessary when developing the general equations of motion of masses in gravitational fields in subsection E.

Finally, the force acting on a mass $m_g$ at an arbitrary point $a$ as seen by an observer at $o$, $\mathbf{F_{ao}}$, is given by

$$\mathbf{F_{ao}} = -\underline{\nabla} E_{pot\_ba}. \tag{37}$$

This equation can either be solved by differentiation with respect to the space coordinates as implied by the gradient or can be put in the form

$$\mathbf{F_{ao}} = \left[\sum_{i=1}^{n}\left(\frac{GM_{gi}m_g(\mathbf{r_i}-\mathbf{r_a})}{|\mathbf{r_i}-\mathbf{r_a}|^3}\right)\right]\exp\left(\sum_{i=1}^{n}\frac{2GM_{gi}}{c^2}\left(\frac{1}{|\mathbf{r_i}|}-\frac{1}{|\mathbf{r_i}-\mathbf{r_a}|}\right)\right) \tag{38}$$

This equation can be proven mathematically using (36) and (37). Note that in the differentiation with respect to space coordinates, the space dilation factor of $\lambda_{oa}$ between $a$ and $o$ should be taken into account by using $ds_a = ds_o/\lambda_{oa}$, where $ds_a$ is the distance at $a$ and $ds_o$ is the distance at $o$. However, (38) has also a physical meaning. In the reference frame of $m_g$, the object sees a space dilation as compared to an observer at $o$ given by $\lambda_{oa}$. This space dilation causes a quadratic scaling of the gravitational force by a factor of $\lambda_{oa}^2$. Hence, the solution in (38) has the traditional Newton's force multiplied by $\lambda_{oa}^2$ represented by the exponential.

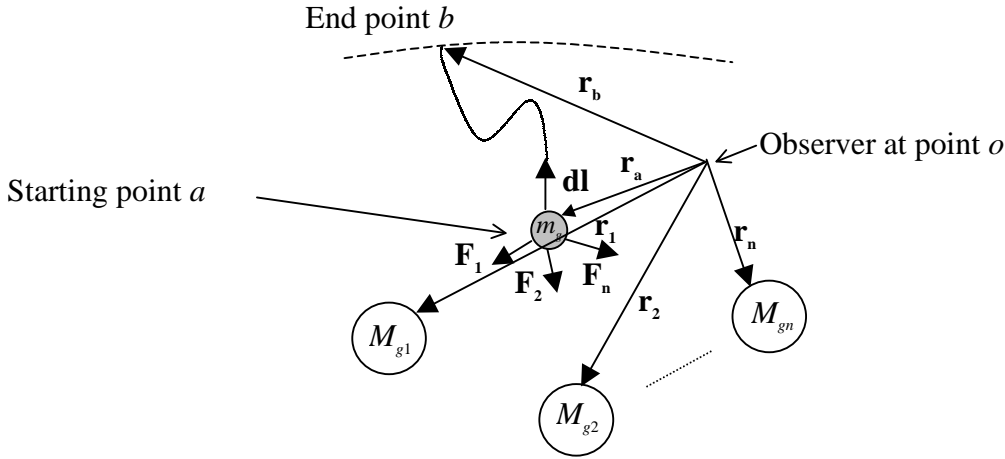

Fig. 7. An object of mass $m_g$ is moved from point $a$ to point $b$ in the gravitational field of a set of static masses $M_{g1}$ - $M_{gn}$. An observer at point $o$ measures the potential energy gained by $m_g$ relative to his frame of reference.

## C. Energy Conservation and Relation to Special Relativity

Consider the two mass problem shown in Fig. 5. An object of mass $m_g$ is moved from a distance $r_1$ to a larger distance $r_2$ in the gravitational field of another mass $M_g$. The object acquires potential energy and is at rest at $r_2$. If the object is left to fall freely under gravity, it gains kinetic energy (represented by speed and motion) while falling to lower potential energy regions. The total energy should remain constant as required by energy conservation, which means that the potential energy lost should be exactly equal to the kinetic energy gained by $m_g$. Hence, if an observer at $r_1$ calculates the potential energy gained by the body when moving to $r_2$ in his frame of reference and observes the speed of the body when it falls back to $r_1$, the observer can find out the form of the kinetic energy conserved by the given potential field. To illustrate this process, assume $m_g$ is moving in a Newtonian gravitational field. The potential energy gained by $m_g$ as seen by the observer at $r_1$, is given by

$$E_{pot21} = GM_g m_g \left( \frac{1}{r_1} - \frac{1}{r_2} \right). \tag{39}$$

To calculate the speed gained by $m_g$ at $r_1$, the force equation can be used and integrated from $r_2$ to $r_1$ as given by

$$\int_0^v v\,dv = -\int_{r_2}^{r_1} G \frac{M_g}{r^2} dr, \quad \text{where} \quad F = m_g \frac{dv}{dt} = -m_g v \frac{dv}{dr}, \tag{40}$$

from which the speed is given by

$$v = \sqrt{2GM_g \left[ \frac{1}{r_1} - \frac{1}{r_2} \right]}. \tag{41}$$

Substituting the factor $GM_g \left[ \frac{1}{r_1} - \frac{1}{r_2} \right]$ from (41) into (39) the potential energy which is transformed into kinetic energy is given by

$$E_{kin} = \frac{1}{2} m_g v^2. \tag{42}$$

Of course, this formula is well known and can be derived in other ways completely independent from the Newtonian gravitational potential. However, if someone does not know what is the form of the kinetic conserved by a Newtonian gravitational potential, he can conclude it as described above.

The same procedure is applied here to find out the form of kinetic energy that is conserved by the gravitational model introduced here. The potential energy gained by $m_g$ when moved from $r_1$ to $r_2$ in the reference frame of an observer at $r_1$ is given by

$$E_{pot21} = m_g c^2 \left[ \exp\left( \frac{GM_g}{c^2} \left( \frac{1}{r_1} - \frac{1}{r_2} \right) \right) - 1 \right]. \tag{43}$$

To find out the speed when the object falls from $r_2$ to $r_1$, it is easier to perform this task in the reference frame of $r_2$ since the moving object maintains the same energy level as that at $r_2$ due to the conservation of energy. Maintaining the same energy level as $r_2$ implies that the moving body maintains the same inertial mass ($m_g$ in the reference frame of $r_2$) and space dilation (see section

II) and hence sees the same forces as an observer at $r_2$. Hence, the speed of the moving body can be calculated from

$$m_g v \frac{dv}{dr} = -G \frac{M_g m_g}{r^2} \exp\left(-2 \frac{GM_g}{c^2}\left(\frac{1}{r} - \frac{1}{r_2}\right)\right) \tag{44}$$

Integrating this equation from $r_2$ to $r_1$ results in a speed of the object when it reaches $r_1$ as observed by an observer at $r_1$ that is given by

$$v = c \sqrt{\left(1 - \exp\left(-2 \frac{GM_g}{c^2}\left(\frac{1}{r_1} - \frac{1}{r_2}\right)\right)\right)}. \tag{45}$$

Note that this speed cannot exceed the speed of light which means that gravitation cannot accelerate (or decelerate) an object to more than the speed of light. The speed of light occurs only at $r_1 = 0$, which requires that the whole of $M_g$ is located inside a zero radius sphere implying an infinite density. This conclusion has significant physical implications but will be discussed in more details in section V.

To continue the procedure of determining what type of kinetic energy is conserved by the gravitational model introduced here, note that the exponential in (43) can be determined from (45) as

$$\exp\left(\frac{GM_g}{c^2}\left(\frac{1}{r_1} - \frac{1}{r_2}\right)\right) = \frac{1}{\sqrt{1 - \frac{v^2}{c^2}}} = \lambda_{21}. \tag{46}$$

Substituting this relation into (43) results in

$$E_{kin} = m_g c^2 \left(\frac{1}{\sqrt{1 - \frac{v^2}{c^2}}} - 1\right). \tag{47}$$

The two above equations are the well-known equations describing the time dilation and kinetic energy in the special relativity theory. This result is a very strong evidence in favor of the gravitational model introduced here since the reader can easily verify that special relativity was not used anywhere in the derivation of the model. Hence, this gravitational model conserves the relativistic kinetic energy instead of the classical kinetic energy conserved by Newtonian gravitational fields. Also, note that the time dilation of the moving object relative to any observer in another reference frame remains the same during a gravitational free fall. This fact is implied by (46). An observer on the moving object will measure his time to be faster than the time at $r_1$ when the object is at rest at $r_2$ due to the higher potential. As the body starts to move towards $r_1$, the observer on the moving object will have two opposite effects on time. He will feel that his time is becoming slower relative to time at $r_1$ due to moving to a lower potential. However, due to its motion at increasing speed, the observer on the moving object will feel that the time at $r_1$ is slowing down. This two opposite effects exactly cancel each other. Equation (46) illustrates that the moving observer will see the time at $r_1$ when he reaches $r_1$ to run slower due to its motion by exactly the same amount he was seeing the time slower at $r_1$ when the object was stationary at $r_2$.

Hence, a general rule can be made that an object moves under gravitational forces such that it maintains its time frame and hence, its view of the world in terms of relative time.

### D. Force on a Body Moving in the Gravitational Field of Another Body

In this section, the equations governing the motion of two masses are discussed. These equations are generalized to a system of any number of masses in subsection E. Consider a mass $m_g$ moving under the gravitational field of another mass $M_g$. The location of $m_g$ relative to $M_g$ is described by the time varying position vector **r** as shown in Fig. 8. The motion of $m_g$ is observed by an observer $O$ at a position described by $\mathbf{r_o}$ relative to $M_g$. Observer $O$ is stationary relative to $M_g$. The corresponding scalar distances from $M_g$ are given by $r$ and $r_o$, respectively. Another stationary observer relative to $M_g$ at distance $r$ (same potential as $m_g$) is $O'$ who observes the motion of $m_g$ in the field of $M_g$. The mass $m_g$ moves with an instantaneous vector velocity given by **v** relative to $M_g$, $O'$, and $O$, as shown in Fig. 8.

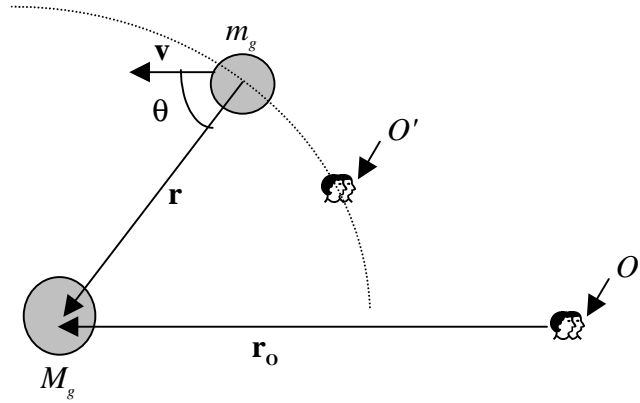

Fig. 8. Calculating the force on an object $m_g$ moving in the gravitational field of another object $M_g$ as observed by an observer $O$.

Determining the motion of $m_g$ in the reference frame of $O$ involves three steps. First, finding the force on $m_g$ as felt by an observer on $m_g$, i.e., in the frame of reference where $m_g$ is stationary and the observer is at the same potential as $m_g$. This step is nontrivial because $M_g$ is in motion relative to $m_g$. The second step is to transform this force to the frame of reference of $O'$. The third step is to transform the force from the frame of reference of $O'$ to the frame of reference of $O$. The first two steps require only the use of special relativity since the observers are at the same potential as $m_g$. The third step requires taking into account the space dilation due to the different potential of $O$ in the same manner used in section II and in determining the potential function.

To find the force on $m_g$, a similar procedure as the one used to determine the force of a moving charge in electrodynamics [46]-[49] is applied. The arguments used in electrodynamics is based only on special relativity [46]-[49] and are completely valid for gravitational fields as in the case of electric fields with the exception of using Newton's law instead of Coulomb's law. Using the procedure described in [46]-[49], the force on $m_g$ as viewed by an observer on $m_g$ is given by

$$\mathbf{F'}_{m_g} = \frac{GM_g m_g}{r'^2} \frac{\left(1 - \frac{v'^2}{c^2}\right)}{\left(1 - \frac{v'^2}{c^2}\sin^2(\theta)\right)^{3/2}} \hat{\mathbf{r}}, \tag{48}$$

where $\theta$ is the angle between the vector $\mathbf{r}$ and the velocity vector $\mathbf{v}$ as shown in Fig. 8. The primes on the variables indicate that these variables are in the reference frame of an observer in the same potential as $m_g$ ($O'$) where special relativity holds. These variables are affected by the space dilation when observed from a different potential. The above equation can be put in vector notation as

$$\mathbf{F'}_{m_g} = \frac{GM_g m_g}{r'^2} \frac{\left(1 - \frac{v'^2}{c^2}\right)}{\left(1 - \frac{v'^2}{c^2}(1 - (\hat{\mathbf{r}} \cdot \hat{\mathbf{v}})^2)\right)^{3/2}} \hat{\mathbf{r}}. \tag{49}$$

The second step is to transform this force to the reference frame of an observer stationary relative to $M_g$ at the same potential as $m_g$, i.e., as observed by $O'$. Again, from special relativity and electrodynamics [1], [3], [4], [46]-[49], the force transforms in the following manner

$$F'_{\parallel} = F'_{m_g \parallel} \quad \text{and} \quad F'_{\perp} = F'_{m_g \perp} \sqrt{1 - \frac{v'^2}{c^2}}, \tag{50}$$

where $\mathbf{F'}$ is the force from the point of view of $O'$ and the parallel and perpendicular subscripts indicate the components of $\mathbf{F'}$ and $\mathbf{F'}_{mg}$ parallel and perpendicular to $\mathbf{v}$, respectively. It is worth noting at this point that in general, the speed used in the above transformation is not the same as the speed used in (49). The speed used here is the relative speed between $O'$ and $m_g$ while the speed used in (49) is the speed of $M_g$ relative to $m_g$. There is no difference between the two speeds in this special case since $O'$ is stationary relative to $M_g$. $\mathbf{F'}$ can be put in vector notation as

$$\mathbf{F'} = m' \frac{d\mathbf{v'}}{dt'} = \frac{GM_g m_g}{r'^2} \frac{\left(1 - \frac{v'^2}{c^2}\right)}{\left(1 - \frac{v'^2}{c^2}(1 - (\hat{\mathbf{r}} \cdot \hat{\mathbf{v}})^2)\right)^{3/2}} \left((\hat{\mathbf{r}} \cdot \hat{\mathbf{v}})\hat{\mathbf{v}} + (\hat{\mathbf{r}} - (\hat{\mathbf{r}} \cdot \hat{\mathbf{v}})\hat{\mathbf{v}})\sqrt{1 - \frac{v'^2}{c^2}}\right) \tag{51}$$

where $m'$ is the inertial mass of $m_g$ at $r$ as opposed to $m$ which is the inertial mass of $m_g$ at $r_o$.

The final step is to transform this equation from the reference frame of $O'$ to the reference frame of $O$ at a different potential. Following the same procedure used in section II, the following transformations hold

$$m' = \lambda_o m, \quad dt' = \lambda_o dt, \quad r' = \frac{r}{\lambda_o}, \quad \text{and} \quad v' = \frac{dr'}{dt'} = \frac{v}{\lambda_o^2}, \tag{52}$$

where

$$\lambda_o = \exp\left(\frac{GM_g}{c^2}\left(\frac{1}{r_o} - \frac{1}{r}\right)\right) \tag{53}$$

Note that $\lambda_o$ is not a constant but is a function of $r$. The variables $v$ and $r$ are the velocity of $m_g$ and distance between $m_g$ and $M_g$ as seen by $O$. The values of $v$ and $r$ are smaller than the values of $v'$ and $r'$ if $r_o$ is larger than $r$ due to the space contraction and time speed up from the point of view of an observer at higher potential. Substituting the transformations from (52) into (51), the following equation describing the motion of $m_g$ as viewed by $O$ and in terms of the space coordinates of $O$ is given by

$$\frac{d\left(\frac{\mathbf{v}}{\lambda_o^2}\right)}{dt} = \frac{GM_g}{r^2}\lambda_o^2 \frac{\left(1 - \frac{v^2}{\lambda_o^4 c^2}\right)}{\left(1 - \frac{v^2}{\lambda_o^4 c^2}(1-(\hat{\mathbf{r}}\cdot\hat{\mathbf{v}})^2)\right)^{3/2}} \left((\hat{\mathbf{r}}\cdot\hat{\mathbf{v}})\hat{\mathbf{v}} + (\hat{\mathbf{r}} - (\hat{\mathbf{r}}\cdot\hat{\mathbf{v}})\hat{\mathbf{v}})\sqrt{1 - \frac{v^2}{\lambda_o^4 c^2}}\right) \tag{54}$$

To illustrate the use of the equation of motion in (54), consider the case when $m_g$ starts with an initial velocity $v_{in}$ in the direction of $\hat{\mathbf{r}}$ at $r = r_2$. The velocity of $m_g$ when it drops to $r = r_1$ is to be determined from the point of view of an observer $O$ at $r_o$. In this case, $\hat{\mathbf{r}} = \hat{\mathbf{v}}$. Hence, (54) becomes

$$\frac{d\left(\frac{v}{\lambda_o^2}\right)}{dt} = \frac{GM_g}{r^2}\lambda_o^2\left(1 - \frac{v^2}{\lambda_o^4 c^2}\right) \tag{55}$$

where $v$ is the velocity in the direction of $\hat{\mathbf{r}}$ and increases as $r$ decreases. By substituting for $dt$ from the relation $dr = -vdt$, the speed at $r_1$ can be determined from

$$\int_{v(r_2)}^{v(r_1)} \frac{(v/\lambda_o^2)}{\left(1 - \frac{v^2}{\lambda_o^4 c^2}\right)} d\left(\frac{v}{\lambda_o^2}\right) = -\int_{r_2}^{r_1} \frac{GM_g}{r^2} dr. \tag{56}$$

This speed is given by

$$v(r_1) = c\lambda_o^2(r_1)\sqrt{1 - \left(1 - \frac{v_{in}^2}{\lambda_o(r_2)^4 c^2}\right)\exp\left(-2\frac{GM_g}{c^2}\left(\frac{1}{r_1} - \frac{1}{r_2}\right)\right)}, \tag{57}$$

where

$$\lambda_o(r_1) = \exp\left(\frac{GM_g}{c^2}\left(\frac{1}{r_o} - \frac{1}{r_1}\right)\right) \quad \text{and} \quad \lambda_o(r_2) = \exp\left(\frac{GM_g}{c^2}\left(\frac{1}{r_o} - \frac{1}{r_2}\right)\right) \tag{58}$$

If $r_o$ is chosen to be $r_1$ (*i.e.*, the observer $O$ is at $r_1$) and $v_{in} = 0$, (57) becomes

$$v(r_1) = c\sqrt{1 - \exp\left(-2\frac{GM_g}{c^2}\left(\frac{1}{r_1} - \frac{1}{r_2}\right)\right)}, \tag{59}$$

which is the same as (45) but derived in a completely different way. As was shown in subsection C, (45) guarantees energy conservation. Also, (45) shows that the maximum speed an observer can see a falling object in the same potential as himself is $c$ (when $r_1 = 0$), which is required since special relativity holds at equipotential regions. An observer at another potential (at $r_o$ less than $r_1$) can however see the object moving with a speed higher than the speed of light in his frame of reference (at $r_o$). However, the speed of the object is always perceived by $O$ as less than the speed of light $c$ in the frame of reference of the moving object (at $r_1$) as illustrated by (57). Note also that if $v_{in} = 0$ at $r_2$, the whole potential energy stored in $m_g$ at $r_2$ relative to $r_1$ from the point of view of $O$ is converted into kinetic energy when the object arrives at $r_1$. This fact can be illustrated by noting that $v(r_1)$ in (57) can be determined by equating the potential energy at $r_2$ relative to $r_1$ and the kinetic energy at $r_1$ as given by

$$m_g c^2 \lambda_o(r_1)\left[\exp\left(\frac{GM_g}{c^2}\left(\frac{1}{r_1} - \frac{1}{r_2}\right)\right) - 1\right] = m_g c^2 \lambda_o(r_1)\left(\frac{1}{\sqrt{1 - \frac{v^2}{\lambda_o^4(r_1)c^2}}} - 1\right) \tag{60}$$

Note that in the above equation, the energy is calculated from the point of view of $O$ who measures energy relative to his frame of reference and hence a factor of $\lambda_o(r_1)$ is multiplied by the potential and kinetic energies from the point of view of an observer at $r_1$. Note also that the speed has to be scaled down by a factor of $\lambda_o^2(r_1)$.

Another interesting case to examine is when an object (such as a photon) starts with the speed of light $c$ at $r_2$ (which appears as $\lambda_o^2(r_2)c$ in the frame of reference of $O$). Substituting this value for $v_{in}$ in (57) results in

$$v(r_1) = c\lambda_o^2(r_1) = c\exp\left(2\frac{GM_g}{c^2}\left(\frac{1}{r_o} - \frac{1}{r_1}\right)\right) \tag{61}$$

The above expression represents the well-known fact that the speed of light changes due to the space and time dilations at different potentials. Light has a peculiar behavior when falling in gravitational fields. While any other object accelerates when falling into a gravitational field, a photon decelerates as it falls into a gravitational field. This behavior is necessary to maintain the constancy of the speed of light despite the space and time dilations which require that light runs slower at lower potentials or closer to the attracting object. Equation (61) is in agreement with this behavior of light because as $r_1$ decreases, $v(r_1)$ decreases. The behavior of light in the presence of a gravitational field can also be described as *gravity repels an object running at the* *speed of light*. This fact is in agreement with equation (55) describing the motion of an object falling in a gravitational field. This force becomes negative when $\dfrac{v^2}{\lambda_o^4 c^2}$ becomes greater than one, which happens when an object runs at a speed higher than the speed of light in the potential frame of reference where the object is. This happens when a photon falls from a higher potential to a lower one since the speed of light in the higher potential is higher than in the lower one. Hence, instantaneously, the photon appears running at a speed higher than the speed of light in the lower potential frame of reference. This creates a repulsive force as discussed before which restores the speed of the photon to the speed of light in the lower potential frame. The above discussion illustrates that the equation of motion in this section is in agreement with energy conservation and the behavior of light in gravitational fields. Later, this equation will also be shown to accurately predict Mercury's precession.

**E. Equations of Motion for a System of Masses**

Consider a set of *n* masses $M_1$-$M_n$ in motion under their gravitational field with certain distribution in space. The objective of this section is to determine the general equations governing the motion of an object $M_{gj}$ at any instant of time *t*, determining its position and speed given the positions and speeds of the ensemble of masses at any instant of time $t_0$. To accomplish this goal, an observation point (or frame) has to be chosen as explained in subsection D since in general, observers at different frames measure different speeds of the masses.

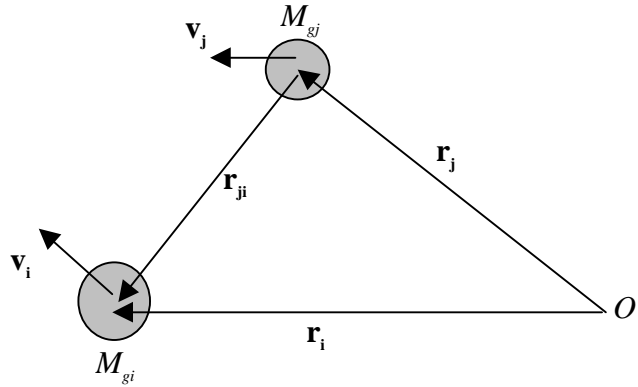

Fig. 9. The force on $M_{gj}$ due to $M_{gi}$ as seen by an observer *O*. There is a total of *n*-1 masses affecting the motion of $M_{gj}$.

Refering to Fig. 9, $M_{gj}$ moves by a speed $\mathbf{v_j}$ while $M_{gi}$ moves by a speed $\mathbf{v_i}$. The vectors from *O* to $M_{gj}$ and $M_{gi}$ are $\mathbf{r_j}$ and $\mathbf{r_i}$, respectively. The distance between the two masses is $\mathbf{r_{ji}}$. With these definitions, the force on $M_{gj}$ due to the *n*-1 masses can be calculated as the superposition of the forces due to each mass. The force due to each mass can be found in the same way as discussed in the previous section. Hence, the equation governing the motion of $M_{gj}$ including the effects of all the masses is given by

$$\frac{d\left(\dfrac{\mathbf{v_j}}{\lambda_{jo}^{2}}\right)}{dt} = \lambda_{jo}^{2} \cdot \sum_{\substack{i=1 \\ i \neq j}}^{n} \left\{ \frac{GM_{gi}}{|\mathbf{r_{ji}}|^{2}} \frac{\left(1 - \dfrac{|\mathbf{v_{ji}}|^{2}}{\lambda_{jo}^{2} c^{2}}\right)}{\left(1 - \dfrac{|\mathbf{v_{ji}}|^{2}}{\lambda_{jo}^{4} c^{2}}(1-(\hat{\mathbf{r}} \cdot \hat{\mathbf{v}}_{ji})^{2})\right)^{3/2}} \left( (\hat{\mathbf{r}}_{ji} \cdot \hat{\mathbf{v}}_{j})\hat{\mathbf{v}}_{j} + (\hat{\mathbf{r}}_{ji} - (\hat{\mathbf{r}}_{ji} \cdot \hat{\mathbf{v}}_{j})\hat{\mathbf{v}}_{j}) \sqrt{1 - \dfrac{|\mathbf{v_j}|^{2}}{\lambda_{jo}^{4} c^{2}}} \right) \right\}. \tag{62}$$

where

$$\lambda_{jo} = \exp\left( \sum_{\substack{i=1 \\ i \neq j}}^{n} \frac{GM_{gi}}{c^{2}} \left( \frac{1}{|\mathbf{r_i}|} - \frac{1}{|\mathbf{r_i} - \mathbf{r_j}|} \right) \right), \tag{63}$$

$$\mathbf{r_{ji}} = \mathbf{r_i} - \mathbf{r_j}, \text{ and } \mathbf{v_{ji}} = \frac{\left(1 - |\mathbf{v_j}|^{2}/c^{2}\right)}{1 - (\mathbf{v_i} \cdot \mathbf{v_j})/c^{2}} (\mathbf{v_i} - (\mathbf{v_i} \cdot \hat{\mathbf{v}}_j)\hat{\mathbf{v}}_j) + \frac{(\mathbf{v_i} \cdot \hat{\mathbf{v}}_j)\hat{\mathbf{v}}_j - \mathbf{v_j}}{1 - (\mathbf{v_i} \cdot \mathbf{v_j})/c^{2}}. \tag{64}$$

Note that the space dilation factor due to potential energy $\lambda_{jo}$ used here is the multi-mass factor described in subsection B. Note also that the speed $\mathbf{v_{ji}}$ is the relative speed between the masses $M_{gi}$ and $M_{gj}$ and is not simply given by $\mathbf{v_j}$-$\mathbf{v_I}$ since the relativistic rule of velocity addition (subtraction) should be employed as given by (64) [1], [3], [4]. With $\mathbf{v_j}$ given by $d\mathbf{r}_j/dt$, the motion of the whole system can be determined by solving $n$ coupled vector differential equations as given by (62) with $j = 1$ to $n$. The $n$ variables solved for are $\mathbf{r_1}$ to $\mathbf{r_n}$. Hence, if the speeds and positions of all the interacting masses in a certain system are recorder at any instant of time $t_0$ as seen from some observation point, the motion of the system as viewed from $O$ can be completely determined.

### IV. Astronomical Calculations Based on the Gravitational Model

In this section, the theory presented in this paper will be employed to calculate the satellite time dilation, the light deflection by the sun, and Mercury's orbit precession. In all these cases, the calculations agree with observations to within 1%. Throughout the section, the calculation steps are explained in a relatively detailed manner. The constants used in these calculations are listed in Table 1.

Table 1. Physical constants used in this section [50], [51]. All constants are in SI units (meter, second, kilogram).

| Constant Description | Symbol | Value |
| --- | --- | --- |
| Sun's mass | $M_{gs}$ | 1.987323 x $10^{30}$ |
| Sun's radius | $R_s$ | 6.9565 x $10^8$ |
| Earth's mass | $M_{ge}$ | 5.98 x $10^{24}$ |
| Earth's radius | $r_e$ | 6.38 x $10^6$ |
| Mercury's orbit semi major axis | $a$ | 5.788 x $10^{10}$ |
| Earth's orbit semi major axis | $a_e$ | 1.495727 x $10^{11}$ |
| Mercury's eccentricity | $e$ | 2.05627 |
| Speed of light in vacuum | $c$ | 2.99 x $10^8$ |
| Newton's gravitational constant | $G$ | 6.668 x $10^{-11}$ |

## A. Satellite Time Dilation

Consider a satellite orbiting the earth in a circular orbit at a distance $b$ from the center of the earth. The speed of the satellite in its orbit can be calculated to a high degree of accuracy using Newtonian mechanics and is given by the well-known formula [50]-[53]

$$v = \sqrt{\frac{GM_{ge}}{b}}. \tag{65}$$

There are three sources of relativistic time dilation between a clock on the surface of the earth and a clock on the satellite. The sources are the gravitational potential difference between the two clocks, the motion of the satellite in its orbit, and the motion of the clock on the earth's surface due to the rotation of the earth. If the clock on the earth is not moving (in the north pole for example) and the satellite was stationary at a height $b$ from the center of the earth, the time dilation is only due to gravitation and is given by equation (23) with $r_1 = r_e$ and $r_2 = b$ which is given by

$$\lambda_1 = \exp\left(\frac{GM_{ge}}{c^2}\left(\frac{1}{r_e} - \frac{1}{b}\right)\right) \tag{66}$$

The above relation represents a speed up in the satellite clock relative to the clock on the earth because $r_e$ is smaller than $b$. However, the satellite is in motion in its orbit with a speed given by (65). Since the stationary satellite at $b$ is in the same potential as the moving satellite, the only source of time dilation between the two satellites is due to motion. This time dilation can be easily found by using the well-known time dilation formula from special relativity and the satellite speed in (65). Hence, the time dilation between the stationary and moving satellites is given by

$$\lambda_2 = \sqrt{1 - \frac{GM_{ge}}{c^2 b}}, \quad (67)$$

which represents a slow down of the clock on the moving satellite relative to the clock on the stationary satellite. Hence, the time dilation between the stationary clock on the earth and the clock on the moving satellite is given by

$$\lambda_3 = \exp\left(\frac{GM_{ge}}{c^2}\left(\frac{1}{r_e} - \frac{1}{b}\right)\right)\sqrt{1 - \frac{GM_{ge}}{c^2 b}}. \quad (68)$$

The earth can be considered approximately as an equipotential surface and the additional time dilation factor due to the motion of the clock on the earth relative to a stationary clock on the earth can be calculated again using special relativity. Hence, the total time dilation between a clock in motion on the earth and the moving satellite is given by

$$\lambda_4 = \exp\left(\frac{GM_{ge}}{c^2}\left(\frac{1}{r_e} - \frac{1}{b}\right)\right)\sqrt{1 - \frac{GM_{ge}}{c^2 b}} \frac{1}{\sqrt{1 - \frac{v_c^2}{c^2}}}, \quad (69)$$

where $v_c$ is the speed of the clock on the earth due to the earth rotation. This speed varies depending on the location of the clock on the earth and ranges from zero at the poles to 463997 m/sec at the equator.

This formula can be applied to the Global Positioning System (GPS) satellites which are positioned at an orbit 4.1 the earth's radius $r_e$ ($b$ = 2615800 m) from the center of the earth [9]-[13]. A clock on such a satellite would run faster than a clock at the equator by a factor of 444.66 parts in $10^{12}$ according to (69) and using the values in Table 1. For a clock in the north pole, the speed up is 443.45 parts in $10^{12}$. These speed-up factors correspond to a 38410 and 38310 ns/day difference between the satellite clock, and the equator and north pole clocks, respectively. An experiment was conducted with the NTS-2 satellite, which contained the first Cesium clock to be placed in orbit [9]-[13]. The atomic clock was operated for about 20 days to measure its clock rate. The frequency measured during that interval was +442.5 parts in $10^{12}$ faster than clocks on the ground [9]-[13]. The predictions by the theory presented here are within 2.2 parts in $10^{12}$ in the worst case (when the clock is in the equator), well within the precision of the orbiting clock [9]-[13]. The relative error between measurement and calculations is less than 0.5% (2.2/442.5*100).

## B. Light Deflection by the Sun

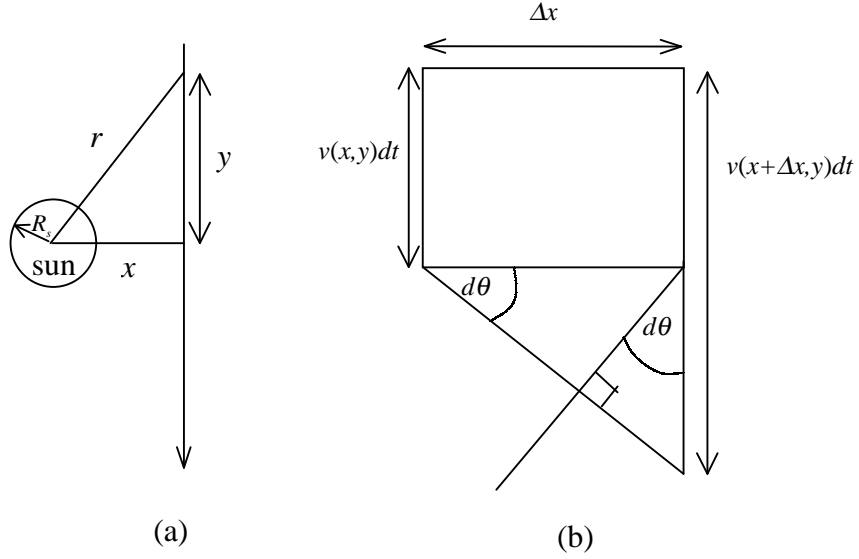

(a)            (b)

Fig. 10. Light deflection by the sun. (a) A ray of light is travelling from $y = -\infty$ to $y = \infty$ at a distance $x$ from the sun. (b) The incremental rotation of the wave front of light due to the speed difference at locations $(x+\Delta x, y)$ and $(x, y)$ relative to the vertical direction.

Consider a ray of light travelling from $y = -\infty$ to $y = \infty$ at a distance $x$ from the sun as shown in Fig. 10 (a). It is required to determine the angle of deflection of this ray of light at $y = \infty$ relative to the vertical direction. Referring to Fig. 10 (b), the wave front of light rotates relative to the vertical direction due to the speed difference at locations $(x+\Delta x, y)$ and $(x, y)$. From Fig. 10 (b), the incremental rotation angle $d\theta$ at a position $(x, y)$ is given by

$$d\theta = \lim_{\Delta x \to 0}\left[\frac{v(x+\Delta x, y) - v(x, y)}{\Delta x}\right]dt = \frac{\partial v(x, y)}{\partial x}dt \tag{70}$$

Using the substitution $dy = -v(x,y)dt$, the above relation becomes

$$d\theta = -\left(\frac{1}{v(x, y)}\frac{\partial v(x, y)}{\partial x}\right)dy \tag{71}$$

The light speed in a gravitational field was calculated in section III D (equation (61)) and is given by

$$v(r) = c\exp\left(2\frac{GM_{gs}}{c^2}\left(\frac{1}{r_o} - \frac{1}{r}\right)\right) = c\exp\left(2\frac{GM_{gs}}{c^2}\left(\frac{1}{r_o} - \frac{1}{\sqrt{x^2 + y^2}}\right)\right) \tag{72}$$

Substituting (72) into (71), the following relation results

$$d\theta = -2\frac{GM_{gs}x}{c^2(x^2 + y^2)^{3/2}}dy. \tag{73}$$

Therefore, the total deflection when the ray travels from $y = -\infty$ to $y = \infty$ is given by

$$\theta = -4 \frac{GM_{gs} x}{c^2} \int_{-\infty}^{0} \frac{1}{(x^2 + y^2)^{3/2}} dy. \tag{74}$$

Using the mathematical substitution

$$u = \frac{1}{\sqrt{x^2 + y^2}}, \tag{75}$$

the above integration transforms to

$$\theta = 4 \frac{GM_{gs}}{c^2} \int_{0}^{1/x} \frac{u}{\sqrt{\frac{1}{x^2} - u^2}} dy. \tag{76}$$

This integration can be easily performed and the resulting light deflection is given by

$$\theta = 4 \frac{GM_{gs}}{c^2 x}. \tag{77}$$

This is the well-known light deflection formula that general relativity produces [14]-[20]. For $x = R_s$, the deflection angle is given by 1.75" (arc seconds) in good agreement with observations [14]-[20].

## C. Mercury's Orbit Precession

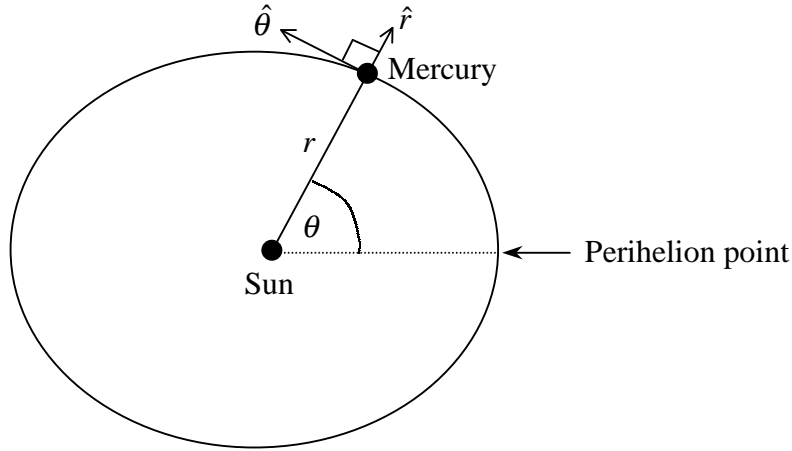

Fig. 11. Mercury's orbit around the sun characterized by polar coordinates.

To calculate Mercury's Perihelion advance, it is convenient to use the polar coordinates as shown in Fig. 11. The acceleration in the polar coordinates in the directions of $\hat{r}$ and $\hat{\theta}$ are given by [50]-[53]

$$\frac{d^2 r}{dt^2} - r\left(\frac{d\theta}{dt}\right)^2 = a_r, \tag{78}$$

$$r\frac{d^2\theta}{dt^2} + 2\left(\frac{dr}{dt}\right)\cdot\left(\frac{d\theta}{dt}\right) = a_\theta, \tag{79}$$

respectively. Note that in Newtonian mechanics, $a_\theta$ is equal to zero since all the force is in the direction of $\hat{r}$. However, under the gravitational model introduced here, $a_\theta$ is not equal to zero since the force has a component in the direction of $\hat{\theta}$ as given by equation (54). To determine $a_r$, the component perpendicular to motion in (54) is utilized which is given by

$$\frac{d\left(\frac{v_r}{\lambda_o^2}\right)}{dt} = -\frac{GM_{gs}}{r^2}\lambda_o^2, \tag{80}$$

where $v_r$ is the velocity component in the direction of $\hat{r}$ and

$$\lambda_o = \exp\left(\frac{GM_{gs}}{c^2}\left(\frac{1}{r_o}-\frac{1}{r}\right)\right)\bigg|_{r_o=\infty} = \exp\left(-\frac{GM_{gs}}{c^2 r}\right) \tag{81}$$

Note that the observer is taken to be at infinity and that the approximation is made that $\hat{r}$ is almost perpendicular to the direction of motion. Substituting for $\lambda_o$ and in (80) and differentiating the left hand side, the following relation results

$$\exp\left(2\frac{GM_{gs}}{c^2 r}\right)\frac{dv_r}{dt} - 2\frac{GM_{gs}}{c^2 r^2}\exp\left(2\frac{GM_{gs}}{c^2 r}\right)v_r = -\frac{GM_{gs}}{r^2}\exp\left(-2\frac{GM_{gs}}{c^2 r}\right) \tag{82}$$

The second term on the left hand side is much smaller than the right hand side since $v_r/c^2 \ll 1$. Hence, the acceleration $a_r$ is given by

$$a_r = \frac{dv_r}{dt} = -\frac{GM_{gs}}{r^2}\exp\left(-4\frac{GM_{gs}}{c^2 r}\right) \tag{83}$$

Hence, equation (78) becomes

$$\frac{d^2 r}{dt^2} - r\left(\frac{d\theta}{dt}\right)^2 = -\frac{GM_{gs}}{r^2}\exp\left(-4\frac{GM_{gs}}{c^2 r}\right) \tag{84}$$

Equation (79) can be put as

$$\frac{1}{r}\frac{d}{dt}\left(r^2\frac{d\theta}{dt}\right) = a_\theta, \tag{85}$$

form which the following relation results

$$r^2\frac{d\theta}{dt} = \int r a_\theta\, dt = h(t). \tag{86}$$

In Newtonian mechanics, $a_\theta = 0$ and $h(t)$ is just a constant. Since $a_\theta$ is not equal to zero in our case, $h(t)$ is a variable function of time. Using (86) to eliminate the time variable from (84) and expressing the derivatives in terms of $1/r$ [50]-[53], the following relation results

$$\frac{d^2(1/r)}{d\theta^2} + \frac{1}{r} = \frac{GM_{gs}}{h(t)^2} \exp\left(-4\frac{GM_{gs}}{c^2 r}\right) \tag{87}$$

To determine $h(t)$, note that in the reference frame of an observer on Mercury (in the same potential and in motion with the object that the force acts on), $a_\theta = 0$ since according to force equation in (48), all the force is in the direction of $\hat{r}$. Hence, the equivalent of (86) in that frame of reference is given by

$$r'^2 \frac{d\theta'}{dt'} = h', \tag{88}$$

where $h'$ is a constant to be determined later and the primes refer to that fact that these variables are in the frame of reference of the moving object. The quantity $h(t)$ can be determined by transforming (88) to the frame of reference of a stationary observer at infinity by using the following transformations (in a similar manner to (52) but taking into account motion as well as potential)

$$dt' = \lambda_o \sqrt{1 - \frac{v^2}{\lambda_o^4 c^2}} dt, \quad r' = \frac{r}{\lambda_o}, \quad dl' = \frac{dl}{\lambda_o \sqrt{1 - \frac{v^2}{\lambda_o^4 c^2}}}, \quad \text{and} \quad d\theta' = \frac{dl'}{r'} = \frac{d\theta}{\sqrt{1 - \frac{v^2}{\lambda_o^4 c^2}}}, \tag{89}$$

where $dl$ is the arc distance cut by the Mercury in its orbit. Note that the time in the reference frame of the moving object appears to an observer at infinity slower due to the lower potential and due to motion. The radial distance $r$ appears smaller only due to the lower potential since $r$ is approximately perpendicular to the direction of $v$. The arc distance $dl$ appears smaller to an observer at infinity due to both potential and motion since this distance is in the direction of $v$. Since the change in the angle $\theta$, $d\theta$, is given by $dl/r$, $d\theta$ appears smaller due to motion only as given by (89). Substituting these transformations into (88), $h(t)$ can be found as

$$r^2 \frac{d\theta}{dt} = h(t) = h' \exp\left(-3\frac{GM_{gs}}{c^2 r}\right)\left(1 - \frac{v^2}{\lambda_o^4 c^2}\right) \tag{90}$$

with $r$ and $v$ as functions of time. Substituting $h(t)$ into (87), the following differential equation between $r$ and $\theta$ results

$$\frac{d^2(1/r)}{d\theta^2} + \frac{1}{r} = \frac{GM_{gs}}{h'^2} \frac{\exp\left(2\frac{GM_{gs}}{c^2 r}\right)}{\left(1 - \frac{v^2}{\lambda_o^4 c^2}\right)^2}. \tag{91}$$

The velocity $v$ can be eliminated from the above equation by considering the total energy (potential + kinetic) of Mercury moving in its orbit. The total energy from the point of view of an observer at infinity can be directly found as

$$m_g c^2 \left[\exp\left(-\frac{GM_g}{c^2 r}\right) - 1\right] + m_g c^2 \exp\left(-\frac{GM_{gs}}{c^2 r}\right)\left(\frac{1}{\sqrt{1 - \frac{v^2}{\lambda_o^4 c^2}}} - 1\right) = U, \quad (92)$$

where $m_g$ is Mercury's gravitational mass and $U$ is the total energy, which is constant due to energy conservation. Hence,

$$\frac{1}{\sqrt{1 - \frac{v^2}{\lambda_o^4 c^2}}} = \left(1 + \frac{U}{m_g c^2}\right) \exp\left(\frac{GM_g}{c^2 r}\right). \quad (93)$$

The value of $U$ is known from Newtonian Mechanics to be equal to $-GM_{gs}m_g/2a$ [50]-[53] where $a$ is the semi major axis of the orbit of Mercury around the sun. Substituting the values in Table 1, the quantity $U/m_g c^2$ is in the order of $10^{-9}$ and can be neglected with respect to one without significantly affecting the accuracy of the calculations. Hence,

$$\frac{1}{\sqrt{1 - \frac{v^2}{\lambda_o^4 c^2}}} = \exp\left(\frac{GM_g}{c^2 r}\right). \quad (94)$$

Substituting this relation into (91), a differential equation that only has $r$ and $\theta$ results

$$\frac{d^2(1/r)}{d\theta^2} + \frac{1}{r} = \frac{GM_{gs}}{h'^2} \exp\left(6\frac{GM_{gs}}{c^2 r}\right). \quad (95)$$

By expanding the exponential into the first two terms and neglecting the third and higher order terms, the following equation results

$$\frac{d^2(1/r)}{d\theta^2} + \omega^2 \frac{1}{r} = \frac{GM_{gs}}{h'^2}, \quad (96)$$

where

$$\omega = \sqrt{1 - 6\left(\frac{GM_{gs}}{ch'}\right)^2}. \quad (97)$$

The solution for (96) is a simple harmonic oscillator given by

$$r = \frac{h'^2}{GM_{gs}} \frac{1}{\left(\frac{1}{\omega^2} + e \cdot \cos(\omega\theta)\right)} \approx \frac{h'^2}{GM_{gs}} \frac{1}{(1 + e \cdot \cos(\omega\theta))}, \quad (98)$$

To find the value of $h'$, the above equation can be compared to the equation describing an elliptical orbit in terms of the semi major axis $a$ and the eccentricity $e$ which is given by

$$r = \frac{a(1-e^2)}{(1 + e \cdot \cos(\omega\theta))}, \quad (99)$$

from which $h'^2 = GM_{gs}a(1-e^2)$. Hence, $\omega$ becomes

$$\omega = \sqrt{1 - 6\left(\frac{GM_{gs}}{c^2 a(1-e^2)}\right)}. \quad (100)$$

From (99), Mercury is at its perihelion (shortest distance from the sun) when $\omega\theta_p = 2n\pi$, where $n$ is an integer taking the values 0, 1, 2, ... Hence, the angle advance in the angle $\theta$ at which the perihelion occurs per revolution is given by $\Delta\theta_p = 2\pi(1/\omega - 1)$. The number of revolutions that Mercury makes per century of the earth's time is given by [50]-[53]

$$revs\_century = 100 \cdot \sqrt{\frac{a_e^3}{a^3}}, \quad (101)$$

where $a_e$ is the semi major axis of the earth. Transforming the precession angle from radians to arc seconds, the precession of Mercury per century in arc seconds is given by

$$precession\_century = 2 \cdot 180 \cdot 3600 \cdot 100 \cdot \sqrt{\frac{a_e^3}{a^3}} \left(\frac{1}{\sqrt{1 - 6\left(\frac{GM_{gs}}{c^2 a(1-e^2)}\right)}} - 1\right). \quad (102)$$

Substituting the values from Table 1 into the above equation gives a precession of 43.188" per century with a relative error of 0.6% as compared to the observed value of 42.95" [21]-[23].

**V. Physical Implications**

The gravitational model introduced here has several significant physical implications. This section briefly points out some of the most significant implications of the new model and how the new concepts agree with physical intuition and observational facts. The section is arranged as follows. In subsection A, it is shown that this model removes all the mathematical infinities and physical puzzles that can arise in general relativity and Newtonian mechanics. In subsection B, the equivalence between the rest energy and potential energy of the universe is shown to flow naturally out of the new model and it is shown that the physical meaning of this observation is that the gravitational potential energy is the source of all other known forces including the electromagnetic forces. In subsection C, signs of compatibility of this model with quantum mechanics are discussed and it is shown that the essence of quantum mechanical behavior is gravitational and that electrons are real black holes.

## A. Removing the Mathematical Problems in General Relativity and Newtonian Mechanics

The theory of general relativity has several mathematical problems that arise in some special cases of very strong gravity. These infinities have been taken by many highly respected scientists (including Einstein himself) as signs of flows in the theory [25]-[27], [32]-[35]. The first physical inconsistency is that the theory of general relativity leads to solutions where time can become imaginary such as in the interior of a black hole where all mathematics break. Besides the mathematical problems, many philosophical dilemmas arise when trying to figure out what happens inside a black hole. Black holes as they are described by the theory of general relativity are one of the greatest puzzles of science and have lead to a huge amount of confusion and dispute, *e.g.,* [29]-[35].

It was shown that according to the theory introduced here, the gravitational time dilation is given by exponential functions (*e.g.*, see (34)). Hence, the value of time dilation is always positive and cannot be imaginary or negative. It was also shown that the gravity can never accelerate or decelerate any object to higher than the speed of light according to (45). An object can be accelerated to exactly the speed of light if there is an infinite mass density at some region of space where all the mass is concentrated in zero radius ($r_1 = 0$). Still, time can never get imaginary under the gravitational model, even at the speed of light. An alternative way to state this characteristic is that light can escape any object of finite mass density. Hence, under this new theory, black holes cannot be objects with finite densities. This characteristic eliminates all the puzzles associated with black holes, imaginary time, and the possibility of travelling to higher than the speed of light under gravitational forces.

By their definition, black holes cannot emit energy or electromagnetic radiation (including light) [26]-[35]. However, non-visible objects were discovered by their influence on other visible stars which emit large amounts of X and gamma radiation in direct violation of the definition of a black hole. The estimated mass of these objects (based on their effects on the other stars) exceeds the Chandrasekhar limit [29], [30]. According to general relativity, if any object exceeds this limit, it has to behave as a black hole and emit no radiation. Hence, these objects illustrate an obvious contradiction between general relativity predictions and observations. Instead of questioning general relativity, complex theories were developed to account for this radiation [27]. Hence, more complexities are added to an already complex model.

These observations can be easily explained based on the theory introduced here. These alleged black holes are objects of extremely high but finite density, which means that they can never trap light and will radiate. There are two opposing effects as an object collapses to such large densities. The radiation from such an object is red shifted due to its high gravity but also the radiation itself is at very high frequency due to the extreme density of the matter inside such an object. Hence, if the increase of the radiation frequency due to higher density is faster than the decrease in frequency due to the red shift, denser objects will radiate at higher frequencies. These dark objects are normal stars that radiate at the X and gamma frequency ranges and hence these stars are not visible in the red to blue frequencies.

Note that it is not claimed here that the above description is a theory. This discussion just points out that there may be much simpler and physically intuitive explanations for these objects under the new theory. Also, note that it is not claimed that black holes do not exist.

Actually, they do exist, but are the most unexpected objects as will be discussed in subsection *C*. The behavior of these black holes is however very different from the behavior of general relativity's black holes and leads to no puzzles.

Another serious infinity that arises in the general relativity theory and Newtonian mechanics, is that the potential energy of an object in reference to $r = 0$ is infinite. That is why the potential energy reference is always taken at infinity in these theories. However, the *physically meaningful* potential energy is the one referred to $r = 0$. This topic will be discussed in the next section.

### B. Equivalence of Rest and Potential Energies (Feynman's Great Mystery)

As mentioned in the introduction, Feynman among many others [26], [27] has pointed out the equivalence between the total potential energy of the universe and the rest energy given by $Mc^2$ with $M$ as the total mass of the universe. This observation cannot be explained using the general relativity theory or Newtonian mechanics. To show why this is so, first the correct potential energy reference should be defined. Consider again the potential energy of a mass $m_g$ in the gravitational field of another mass $M_g$. As long as there is nonzero distance between the two objects, the objects still have potential energy since they can still move and fall on each other if left on their own. If the objects have nonzero radius and touch each other, the potential energy appears as pressure between the two objects. Hence, the objects have some energy to use unless the two objects are stationary at zero distance from each other. That is why the real potential energy should be calculated in reference to $r = 0$ between the two objects where the potential energy is really zero and there is no more energy to spend. If this is attempted in Newtonian mechanics, the potential energy of any body is infinite since the Newtonian potential is given by

$$E_{pot\_ab} = -GM_g m_g \left( \frac{1}{r_a} - \frac{1}{r_b} \right). \tag{103}$$

Hence, a body has to lose an infinite amount of energy to reach $r_a = 0$, although the total energy of the object is given by $m_g c^2$. Even at a finite $r_a$, a body can lose more than its total energy as it falls from $r_b$ to $r_a$. The condition is very similar to the condition for having a black hole, namely when

$$\left[ \frac{E_{pot\_ab}}{m_g c^2} = -\frac{GM_g}{c^2} \left( \frac{1}{r_a} - \frac{1}{r_b} \right) \right] < -1. \tag{104}$$

The same behavior occurs in general relativity. This behavior proves non-physical and very problematic when dealing with elementary point particles such as photons and electrons in quantum mechanics. Also, this behavior is not compatible with theories that assume that the space is filled with particles and antiparticles since this would imply infinite energy[24]-[27].

However, the total potential energy in the universe was calculated by many scientists in reference to $r = 0$. The idea is to model the universe as a sphere with uniform density. Under this model, Newton's law allows the calculation of the potential energy of the mass in the sphere in reference to $r = 0$, which represents the real potential energy. Performing this task requires calculating a double integral using Newton's law. This calculation has been repeated several times using the best known estimates of the universe size and density only to confirm more accurately the relation between the rest mass and potential energy of the universe [25]-[27]. This is now a well-accepted fact but with no viable explanation [25]-[27].

To illustrate how the new model naturally explains this observation, consider again the potential energy of a mass $m_g$ in the gravitational field of another mass $M_g$. The equation describing the potential energy lost when moving $m_g$ from a larger radius $r_b$ to a smaller one $r_a$ as seen by an observer at $r_b$ was shown in section III to be given by

$$E_{pot\_ab} = -m_g c^2 \left[ 1 - \exp\left( -\frac{GM_g}{c^2}\left(\frac{1}{r_a} - \frac{1}{r_b}\right)\right)\right]. \tag{105}$$

It is trivial to check that an observer at $r_b$ will see that $m_g$ lost exactly its whole energy given by $m_g c^2$ as it falls to $r_a = 0$. Note that $M_g$ also loses its whole energy in this process since from its perspective it has fell to $r_a = 0$ in the field of $m_g$. Hence, the total potential energy of the system of masses is equal to the total rest energy of the masses. Note that $m_g$ and $M_g$ will also lose all their energy if brought in rest to zero distance even if there are other masses in the universe. This can be easily checked form the multiple mass potential energy equation given by

$$E_{pot\_ab} = -m_g c^2 \left[ 1 - \exp\left( -\sum_{i=1}^{n} \frac{GM_{gi}}{c^2}\left(\frac{1}{r_{ia}} - \frac{1}{r_{ib}}\right)\right)\right]. \tag{106}$$

as described in section III. Hence, if a particle or mass is forced to zero distance at rest from another particle or mass, both masses lose their full potential energy. It can be also easily verified that if the whole universe is brought in rest to zero radius, the universe losses all its rest energy. This discussion clearly shows that it is inherent in the new gravitational model that the potential energy of any system of particles or masses is equal to their rest energy. Note that it is not required here to model the universe as a sphere of uniform density. The universe can be made of arbitrary discrete and uniform mass distributions and the equivalence between the potential energy and rest energy still holds.

The physical meaning of this equivalence between rest and potential energies can be clarified by examining the behavior of a hydrogen atom in a gravitational field. As a hydrogen atom approaches a mass, it losses potential energy. As discussed in section II, losing this energy means less electron mass, which in turn results in expanding the orbit of the electron around the nucleus. This size expansion is canceled by an equivalent space dilation as discussed before. Hence, the net result that an observer at higher potential will see is that as the hydrogen atom moves towards the other mass, it keeps its size, but all the forces become weaker which results in slowing all the physical processes relative to the higher observer. As the hydrogen atom approaches $r = 0$ from the attracting object, it losses all the forces and energies within it from the perspective of the higher observer. Hence, all the energy in the universe is gravitational potential energy and this energy translates into all other types of energies such as electromagnetic energy and atomic binding energy. That is why the potential energy of any individual body in the universe is equal to the rest energy of that body. This rest energy represents all the internal forces and energies within that body. *Hence, gravitation is the source of all other forces and energies.* Also, gravitational mass is the essential characteristic that any particle has to have to exist. Particles can have no charge or no rest mass, but always have to have a gravitational mass [5]-[8].

## C. Relations to Quantum Mechanical Behavior

The gravitational theory introduced here agrees with the well-known quantum mechanical concepts and facts. Actually, it is shown in this section that gravity is a primary force governing the quantum mechanical behavior of elementary particles. To start this discussion, consider bringing two particles together somehow to rest at zero distance. Theoretically, this is not impossible since unlike Newtonian mechanics and general relativity, this process does not require infinite energy in the new theory. This process requires energy equal to the rest energies of the two particles. As discussed in the previous section the two particles lose all their rest energy as the particles reach zero distance from each other. Assume that each particle has a gravitational mass of $m_g/2$. This particle will have a gravitational mass of $m_g$ since the gravitational mass is always conserved. However, the rest inertial mass of this particle will be zero. *In general, any uncharged point particle will have zero rest inertial mass according to this theory.* Hence, under the slightest gravitational force, this particle will experience infinite acceleration since its inertial mass is zero while its gravitational mass is finite. This finite gravitational mass results in a gravitational force on a particle that has no resistance to motion. As the velocity of this particle increases, its inertial mass changes according to the formula

$$m_{inertial} = \frac{m_{rest}}{\sqrt{1 - \frac{v^2}{c^2}}}. \tag{107}$$

The inertial mass will remain zero at all speeds except at the speed of light $c$, in which case the value $m_{inertial}$ becomes undetermined and can be nonzero. The gravitational theory introduced here is not sufficient alone to determine what $m_{inertial}$ will be at the speed of light and extra information about the physical nature of the problem and the energy exchange involved is needed to determine the inertial mass. Also, once the particle reaches the speed of light, it maintains the speed of light as discussed in section III D.

The gravitational theory introduced here predicts many interesting characteristics for this particle. This particle is a point particle with no dimensions (zero radius) and has to travel at the speed of light as long as there is gravity. Even if the particle can be put to halt by an infinitely accurate gravitational equilibrium condition, this condition is unstable, and the slightest fluctuation will cause this particle to fly at the speed of light. The time it takes this particle to go from rest to the speed of light is zero due to the infinite acceleration. Hence, such a particle can only be found running at the speed of light and is a point particle. This description fits well the behavior of a photon. Hence, starting from the gravitational theory, we can predict the existence of photons. As for the inertial mass of a photon at the speed of light, it is well known from quantum mechanics to be equal to $hf/c^2$ for a photon of frequency $f$. The gravitational mass of a photon is an open question, but it has to be proportional to frequency since all light bends by the same amount under gravitation, which requires a constant ratio of gravitational and inertial mass at all frequencies. Together with the zero rest mass property and the zero dimension property, these set of parameters completely determines the behavior of any photon.

Another interesting conclusion can be made by trying to answer the question of what are the characteristics of a particle that can trap a photon based on the new gravitational theory. As discussed before, an object going at the speed of light can escape any object, except if the object has zero radius. In that case, if a photon falls into that object, it cannot escape, since the photon falls on it with the speed of light which is exactly the speed of escape. In fact, it is more accurate to say that that it is undetermined if the photon will be trapped or not by this object since the speed of light is exactly equal to the escape speed from this object. In any case, an object that can trap a photon has one definite characteristic. It has to be a point particle with zero dimensions according to the new gravitational theory. In effect, the object should be a black hole to trap light. Electrons are known to trap photons (the common term used in quantum mechanics is absorb). Hence, this theory also predicts that an electron has to be a point particle with zero dimensions. All the experiments made to measure the radius of an electron or a photon failed [27], [47]-[49]. It is now fairly well-accepted fact of quantum mechanics that photons and electrons are point particles [27], [47]-[49]. Note that according to general relativity, an electron is not required to be of zero dimensions to trap light since the theory of general relativity allows black holes of nonzero dimensions.

As aforementioned, it is undetermined whether a photon falling on an electron will be trapped or not since the speed of the photon is exactly equal to the escape speed form the electron. If the escape speed from a black hole were just a little larger than that of light, then *any* photon falling onto *any* electron would be trapped. It is well known that this is not how electrons and photons interact. In some cases an electron traps a photon (if the energy of the photon exactly matches the band gaps in the atom) and in the overwhelming majority of cases the photon escapes. The conditions that determine if a photon is trapped or not requires further investigations. However, the new theory introduced here allows both cases. Note again that general relativity would predict that any electron would trap any photon since a black hole in general relativity have regions beyond the event horizon where the escape velocity is greater than that of light.

Also, another consistency between quantum mechanical observations and the theory introduced here is the quantum leaps an electron makes between different orbits of an atom. It is well-known that an electron bound to an atom does not move in a continuous manner between orbits. It rather changes orbits in quantum leaps. This fact can be understood by recalling that a trapped photon moving under the speed of light, once released will experience infinite acceleration to the speed of light. Hence, the photon's energy changes from zero to *hf* in zero time. The initial zero energy of a photon is due to its zero inertial mass below the speed of light as discussed before. This violent change of energy and speed results in an infinite force, which by Newton's third law, results in an infinite reaction force on the electron until the electron reaches the other orbit. Hence, if both momentum and energy conservation are considered together under the new theory, the quantum leaps of an electron between atom orbits can be easily understood.

Therefore, it was illustrated in this section that the force governing the behavior of photons and the interaction between photons and electrons is gravity. This characteristic challenges many of the established perceptions. It is traditionally believed that quantum mechanics and gravitation does not match. However, the problem is not in gravitation but in general relativity, which fails with point particles due to the infinite potentials that general relativity predicts. Also, quantum mechanics is usually classified as relativistic and non-relativistic. According to this discussion, quantum mechanics in all its forms is just extreme relativity. Once the facts are put together, it just appears natural that gravity controls the behavior of photons. After all, a photon is only affected by gravity and mechanical collisions. Electromagnetic forces for example have no effect on photons since photons carry no charge.

## VI. Where Did General Relativity Go Wrong

The answer to this question is clear from the previous discussion. The general relativity depends fundamentally on the equivalence principal between the inertial and gravitational masses. It was shown in this paper that the principle of equivalence has limited validity. It is only true if an experiment and observations are made in the same frame of reference but is not true when observations to an experiment are made from a different frame of reference. However, general relativity works in some cases such as the calculation of Mercury's orbit and calculating the time dilation of a satellite. The reason is that general relativity works at low gravity but fails at strong gravitation [25], [32]-[35]. Intuitively, if the gravity is weak, assuming that the gravitational mass is equal to the inertial mass is valid as a first order approximation. Note that when moving an object from a lower potential to a higher potential, the gravitational mass remains constant while the inertial mass increases by an amount equal to the potential energy over $c^2$. If the gravitation is low, the difference can be neglected. This fact can be illustrated using a specific example. Consider a mass $m_g$ that is moved from a distance $r_1$ from earth to a higher distance $r_2$. Since the earth gravitation is weak, Newton's law can be used as a good approximation. In that case the potential energy difference is given by

$$E_{pot} = Gm_g M_g \left[ \frac{1}{r_1} - \frac{1}{r_2} \right]. \tag{108}$$

Hence, according to the theory introduced here, the time dilation between $r_1$ and $r_2$ is given by

$$\lambda = 1 + \frac{E_{pot}}{m_g c^2}. \tag{109}$$

According to the principle of equivalence, acceleration is equivalent to gravitation. Hence, a body accelerated by the same acceleration as the gravity of the earth between $r_1$ and $r_2$, reaches a velocity given by

$$v = \sqrt{2GM_g \left[ \frac{1}{r_1} - \frac{1}{r_2} \right]}. \tag{110}$$

From special relativity, the time dilation at that speed is given by

$$\lambda = \frac{1}{\sqrt{1 - \frac{v^2}{c^2}}}. \tag{111}$$

Using (108) and (110), the time dilation can be expressed as [26]

$$\lambda = \frac{1}{\sqrt{1 - \frac{2E_{pot}}{m_g c^2}}} \approx 1 + \frac{E_{pot}}{m_g c^2}. \tag{112}$$

The time dilations according to the theory introduced here in (109) and according to general relativity are approximately the same when the potential energy is small as compared to the rest energy of the body but are completely different when these two energies become comparable.

Hence, general relativity appears to make reasonable predictions at low gravitation but fails at extreme gravity, such as in the case of black holes and point particles in quantum mechanics.

A fundamental error was made when deriving the general relativity theory by taking the link between special relativity and general relativity to be acceleration. It can be shown that the real link is energy not acceleration. For example, the time dilation formula in special relativity can be put as

$$\lambda = \frac{1}{\sqrt{1-\frac{v^2}{c^2}}} = 1 + \frac{E_{kin}}{m_g c^2} \quad \text{where} \quad E_{kin} = m_g c^2 \left[\frac{1}{\sqrt{1-\frac{v^2}{c^2}}} - 1\right]. \tag{113}$$

Note that the formula of time dilation is exactly the same in terms of energy both in the gravitational theory introduced here and in special relativity. The space dilation can also be shown to have the same dependence on both kinetic and potential energies. Note that energy is by nature a relative quantity. An absolute value of energy is meaningless. Only the relative values of energy are important and energy has to be measured in reference to another energy level. It is this relativity of energy that underlies the special relativity in case of kinetic energy and as shown here gravitational relativity based on the potential energy. Hence, it is physically intuitive that the link between general and special relativity is energy.

**VII. Conclusions**

A new relativistic gravitational model has been introduced in this paper. Many evidences were provided that support this model such as: the model's self consistency, reduction to Newton's law at low gravitation, the accurate estimation of satellite time dilation, light deflection by the sun, and Mercury's precession, removing the infinities in general relativity, special relativity theory deduction from the new model, agreement with quantum mechanics, and explaining the equivalence between the rest and potential energies. The model was not derived by collecting this set of observations and requirements and trying to fit a theory to them. Rather, the model was derived from a completely different perspective by examining the behavior of an atom in a gravitational field and maintaining energy conservation and the principle of relativity. The agreement of the model developed here with all the known requirements of a correct gravitational model and observational facts is an evidence in favor of the new theory. Also, it was pointed out where general relativity went wrong and that the principle of has limited validity.

Many physical implications follow from the new gravitational model. One implication is that a black hole as defined and characterized by general relativity is wrong. Only an object of infinite density (a point object) can trap light. Also, it was shown that under the new theory an object cannot be accelerated to higher than the speed of light by gravity which is physically intuitive and in agreement with special relativity. It is also shown that time dilation under the new theory has to be positive and cannot be negative or imaginary. The speed of light limit and positive time dilation characteristics seem to be strict rules of nature. The equivalence between rest and potential energies was shown to be in agreement with the new theory means that gravitational potential energy is the source of all energies and forces in the universe. The potential energy simply translates to other forces and fields as discussed in section II.

However, the greatest area to test the new theory is in quantum mechanics and optics. It was shown that the force determining the behavior and interaction between photons and electrons is gravity in its extreme when gravity is so strong to trap light. It was shown that electrons are real black holes (but charged ones). Two mergers between different theories result from this study. Gravitation and quantum mechanics, which are traditionally thought of as non-compatible theories, are actually shown here to be fundamentally related. Relativity and quantum mechanics are also shown to be fundamentally related, with the very basic concepts of quantum mechanics (such as electron-photon interaction) being extreme relativistic effects. Hence, two important missing links are established by the new theory.